\documentclass[prd,10pt,aps,showkeys,twocolumn,notitlepage,longbibliography]{revtex4-2}

\usepackage[separate-uncertainty = true]{siunitx}
\usepackage[utf8]{inputenc}
\usepackage{amsmath}
\usepackage{amsfonts}
\usepackage{amssymb}
\usepackage{microtype}

\usepackage{hyperref}

\usepackage{physics}
\usepackage{mathtools}
\usepackage{gnuplottex}

\usepackage{epstopdf}
\usepackage{tikz,tikz-3dplot}
\usetikzlibrary{3d,patterns}
\tdplotsetmaincoords{50}{120}

\newcommand{\lm}{\ell_m}
\newcommand{\tol}{\mathtt{tol}}
\newcommand{\trp}[1]{#1^{\raisebox{2pt}{\ensuremath{\scriptstyle\intercal}}}}
\newcommand{\iu}{\mathrm{i}}
\newcommand{\entropy}{\mathcal{S}}
\newcommand{\area}{\mathcal{A}}
\newcommand{\field}{\Phi}
\newcommand{\momt}{\Pi}
\newcommand{\pplr}{\chi}
\newcommand{\plr}{\theta}
\newcommand{\azm}{\alpha}
\newcommand{\one}{\ell}
\newcommand{\two}{\mu}
\newcommand{\overbar}[1]{\mkern 2mu\overline{\mkern-2mu#1\mkern-2mu}\mkern 1.5mu}

\DeclareMathOperator{\diag}{diag}

\begin{document}

\title{Universal terms of the entanglement entropy in a static closed universe}

\author{R. R. \surname{Soldati}}
\email{rsoldati@usp.br}\affiliation{Instituto de F\'isica --- Universidade de S\~ao Paulo, CP 66318, 05315-970, S\~ao Paulo-SP, Brazil}
\affiliation{Departamento de F\'isica --- ICEx, Universidade Federal de Minas Gerais, CP 702, 30161-970, Belo Horizonte - MG, Brazil}
\affiliation{Institute for Theoretical Physics I, University of Stuttgart, D-70550 Stuttgart, Germany}
\author{L. S. Menicucci}
\affiliation{Departamento de F\'isica --- ICEx, Universidade Federal de Minas Gerais, CP 702, 30161-970, Belo Horizonte-MG, Brazil}
\author{N. Yokomizo}
\email{yokomizo@fisica.ufmg.br}
\affiliation{Departamento de F\'isica --- ICEx, Universidade Federal de Minas Gerais, CP 702, 30161-970, Belo Horizonte-MG, Brazil}

\begin{abstract}
Subdominant contributions to the entanglement entropy of quantum fields include logarithmic corrections to the area law characterized by universal coefficients that are independent of the ultraviolet regulator and capture detailed information on the geometry around the entangling surface. We determine two universal coefficients of the entanglement entropy for a massive scalar field in a static closed universe $\mathbb{R} \times \mathbb{S}^3$ perturbatively and verify the results numerically. The first coefficient describes a well known generic correction to the area law independent of the geometry of the entangling surface and background. The second coefficient describes a curvature-dependent universal term with a nontrivial dependence on the intrinsic and extrinsic geometries of the entangling surface and curvature of the background. The numerical calculations confirm the analytical results to a high accuracy. The first and second universal coefficients are determined numerically with a relative error with respect to the analytical values of the orders $10^{-4}$ and $10^{-2}$, respectively.

\end{abstract}


\maketitle

% -------------------------------------------------------------

\section{Introduction} \label{sec:intro}

Correlations of vacuum fluctuations of a quantum field in a curved background carry information on the geometry of the background. A key example of the interrelation between the entanglement of field fluctuations and geometry is the celebrated area law \cite{bombelli_quantum_1986,srednicki_entropy_1993}, which states that the dominant contribution to the entanglement entropy of the vacuum for any finite region $A$ of space is proportional to the area $\area$ of its boundary $\Sigma=\partial A$,
\begin{equation}	\label{eq:area-law}
\entropy = \frac{c_2}{\varepsilon^{d-2}} \area + \dots \, ,
\end{equation}
where $\varepsilon$ is an ultraviolet regulator and $d$ is the dimensionality of spacetime. This geometric entropy was originally proposed as a source of entropy for black holes in \cite{bombelli_quantum_1986,srednicki_entropy_1993}. Since then, several works have been devoted to clarifying its role in the quantum mechanics of spacetime, from the formulation of the holographic principle \cite{bousso_holographic_2002} and the description of geometric entropies through the AdS/CFT correspondence \cite{ryu-takayanagi} to the the analysis of emergent properties of spacetime from quantum information-theoretic properties of quantum fields \cite{jacobson_1995,raamsdonk_building_2010,bianchi_architecture_2014,cao_space_2017,saravani_spacetime_2016}.

The fact that properties of the background geometry are imprinted in the field correlations allows one to extract information on the geometry from the network of correlations among the entangled spatial subsystems. In this regard, the area law \eqref{eq:area-law} suffers from the drawback that, in general, it depends on the choice of regularization. However, subdominant terms in the entropy formula \eqref{eq:area-law} include logarithmic divergences that are expect to be regularization independent \cite{Lewkowycz:2012qr}. These can be used to reliably extract geometric information from the entropy function, and provide a natural tool to study the relation between entanglement and geometry.

Universal contributions for the entropy of massive fields were first obtained in \cite{Hertzberg:2010uv}, where the case of a flat waveguide background geometry was considered. For a free massive scalar field and an even number of spacetime dimensions $d$, a contribution of the form
\begin{equation}	\label{eq:universal-1}
\alpha_1(d) \, m^{d-2} \log (m \varepsilon) \area
\end{equation}
was identified, with a universal coefficient $\alpha_1(d)$ that depends only on the dimension $d$. In four dimension, $\alpha_1(d=4)=1/(24 \pi)$. In \cite{Lewkowycz:2012qr}, a curved spherical waveguide background was considered, leading to the identification of new curvature-dependent universal terms. For an even $d \geq 4$ and a spherical entangling surface, a universal term was found of the form
\begin{equation}	\label{eq:universal-2}
\alpha_2(d) \, m^{d-4} \log (m \varepsilon) \area \, ,
\end{equation}
where $\alpha_2(d)$ depends on the Ricci scalar of the intrinsic metric on the entangling surface $\Sigma$ and on the coupling constant $\xi$ that describes the strength of the interaction of the field with the scalar curvature of the background. This universal term describes how the the intrinsic curvature at $(d-2)$-dimensional surfaces $\Sigma$ manifests itself in the entanglement of the field. 

In more general geometries, the universal coefficients $\alpha_i$ are expected to capture more detailed information on the curvature around the entangling surface. Similarly to what happens for conformal field theories \cite{Solodukhin_2008}, one may expect the universal coefficients to depend on several scalars constructed from the the background curvature tensor and the intrinsic and extrinsic curvatures of the entangling surface, as discussed, for instance, in \cite{Lewkowycz:2012qr}. A strategy for computing universal terms of massive theories is provided by the perturbative approach introduced in \cite{Rosenhaus:2014woa,Rosenhaus:2014nha}. This approach was applied to a scalar field in de Sitter space in \cite{Ben-Ami2015}, and indeed led to the discovery of curvature-dependent universal terms with a more complex dependence on the curvature tensors.

The analytical result for the universal term of a conformal field theory in four dimensions obtained in \cite{Solodukhin_2008} was verified numerically for a massless scalar field and spherical entangling surfaces in \cite{Lohmayer:2010}, where the numerical approach originally explored for establishing the area law \cite{srednicki_entropy_1993} was improved to allow for the determination of subleading corrections. The same universal term was derived by different methods in \cite{Casini:2010,Dowker_2010}. The analogous curvature-dependent universal terms for massive theories obtained in \cite{Ben-Ami2015} have not been verified numerically, however, or confirmed by independent alternative derivations.

In this work, we determine universal terms of the entanglement entropy of a massive scalar field for spheres in the Einstein universe $\mathbb{S}^3 \times \mathbb{R}$, both analytically, by the application of the perturbative approach introduced in \cite{Rosenhaus:2014woa,Rosenhaus:2014nha}, and numerically, through the application of the real-time approach originally introduced in \cite{srednicki_entropy_1993} and discussed in the reviews \cite{casini_entanglement_2009,Nishioka:2018khk}. The background geometry in the vicinity of a spherical entangling surface in the Einstein universe is a perturbed spherical waveguide, so that the results of \cite{Lewkowycz:2012qr} describe the zeroth-order term in the perturbations series. The lowest nonzero order terms describe new universal terms of the form \eqref{eq:universal-2}, but with a universal coefficient $\alpha_2(d)$ that depends nontrivially on several scalars constructed from the intrinsic, extrinsic and background curvatures at the entangling surface. 

Being spatially finite, the Einstein universe has a natural infrared cutoff at the scale of its spatial radius, which is convenient for the numerical calculations. In addition, by considering spherical entangling surfaces of distinct radii, the intrinsic and extrinsic curvatures can be varied at the entangling surface, allowing the dependence of the universal coefficients in the distinct curvature terms to be analyzed. As the Einstein universe is static, there is no question as to the choice of the vacuum state, which is unique. Moreover, the discretization required for the numerical calculations can be implemented in a time-independent manner. These properties single out the Einstein universe as a specially convenient geometric background for the study of curvature-dependent universal terms and, in particular, for a numerical test of the analytical techniques employed for their determination.

In Section \ref{sec:model}, we describe the relevant features of the theory of a massive neutral scalar field in the Einstein universe and its discretization. In Section \ref{sec:method}, we briefly review the perturbative approach for the calculation of the universal terms of the entanglement entropy and then apply it to the case of spheres in the Einstein universe. Next, we describe the techniques employed for the numerical calculation of these terms. The numerical results are presented in Section \ref{sec:results}. We summarize and discuss our results in Section \ref{sec:discussion}.

% ------------------------------------------------------------------------------

\section{The model} \label{sec:model}

\subsection{Massive scalar field in the Einstein universe}

The metric of the Einstein universe $ \mathbb{M} = \mathbb{R} \times \mathbb{S}^3 $ in spherical coordinates reads
\begin{equation} \label{eq:line}
	\dd s^2 = -\dd t^2 + R^2 \left( \dd \pplr^2 + \sin^2\pplr \dd \Omega^2 \right) \, ,
\end{equation}
where $\dd \Omega^2 = \dd\plr^2 + \sin^2 \plr \, \dd\azm^2 $ is the metric of the unit $2$-sphere, the coordinates are defined on the intervals $t \in \mathbb{R}$, $\pplr,\plr \in [0,\pi]$ and $\azm \in [0,2\pi]$, and $R$ is the constant radius of the $3$-spheres describing spatial sections at fixed time. The volume element is
\begin{equation}
\sqrt{-g} = R^3 \sin^2 \pplr \sin \plr \, .
\end{equation}
The geodesic distance between antipodal points on the $3$-spheres of constant time is $R\pi$. The area of a spherical surface of fixed $\chi$ is given by
\begin{equation} \label{eq:area}
	\area = 4 \pi R^2 \sin^2 \pplr \, .
\end{equation}

We consider a real massive scalar field $\field(x)$ on this background with a generic coupling to the scalar curvature. The action for the theory in the continuum is
\begin{align*}
I &= \int \dd^4 x  \, \frac{\sqrt{-g}}{2} \left( -\partial_\mu \field \, \partial^\mu \field - m^2 \field^2 - \xi \frac{6}{R^2} \field^2 \right) \\
& \equiv \int \dd{t} L \, ,
\end{align*}
where the last term describes the interaction of the field with the scalar curvature $ 6 / R^2 $ of the metric \eqref{eq:line}. The case of $\xi=0$ describes the minimally coupled theory, while $\xi=1/6$ corresponds to a conformal coupling. After an integration by parts in the angular coordinates $\plr,\azm$, the Lagrangian assumes the form:
\begin{multline}
L = \frac{1}{2} \int_{S^3} \dd{\pplr} \dd{\plr} \dd{\azm} \\ \times \left\{ R^3 \sin^2 \pplr \sin \plr \left[ \dot{\field}^2 - \left(m^2 + \frac{6\xi}{R^2}\right) \field^2 \right] \right. \\
\left.  - R \sin \plr \left[ \sin^2 \pplr (\partial_\pplr \field)^2 -\field \Delta_{ \mathbb{S}^2 } \field   \right] \right\} \, ,
\end{multline}
where
\begin{equation}
\Delta_{ \mathbb{S}^2 }  = \frac{1}{\sin \plr} \partial_\plr (\sin \plr \partial_\plr) + \frac{1}{\sin^2 \plr} \partial_\azm^2 
\end{equation}
is the Laplace-Beltrami operator on the unit $2$-sphere. The momentum associated with the field is the scalar density
\begin{equation}
\momt = \frac{\delta L}{\delta \dot{\field}} = R^3\sin^2 \pplr \sin \plr \,  \dot{\field}  \, ,
\end{equation}
and the Hamiltonian is obtained as usual through the application of a Legendre transformation,
\begin{multline}
H_{ \mathbb{S}^3 } = \frac{1}{2} \int_{S^3} \dd{\pplr} \dd{\plr} \dd{\azm} \, \left\{ \frac{\momt^2}{R^3 \sin^2 \pplr \sin \plr} \right. \\ 
	+ R \sin \plr \left[ \sin^2 \pplr (\partial_\pplr \field)^2 - \field \Delta_{ \mathbb{S}^2 } \field \right]  \\ 
	\left. + R^3 \sin^2 \pplr \sin \plr \left(m^2+\frac{6\xi}{R^2}  \right) \field^2 \right\} \, .
\end{multline}
The canonical fields satisfy the usual Poisson brackets,
\begin{align}
\{\field(x),\momt(x')\} &= \delta(x-x') \, , \nonumber \\
 \{\field(x),\field(x')\} &= \{\momt(x),\momt(x')\} = 0 \, .
\end{align}

The analogue of the momentum representation in spacetimes with spherical spatial sections is obtained by expanding the field in real spherical harmonics,
\begin{equation}
\field(x) = \sum_{\ell = 0}^\infty \sum_{\mu = -\one}^\one  \field_{\ell \mu}(\pplr) Y_{\ell \mu}(\plr,\azm) \, .
\end{equation}
Integrating on the angular variables and using the orthogonality of the spherical harmonics, we obtain the following representation for the Lagrangian:
\begin{multline} \label{eq:L-ell-mu}
L = \sum_{\ell \mu} \frac{R^3}{2} \int_0^{\pi} \dd\pplr \sin^2\pplr \left\{  \left[ \dot{\field}_{\ell \mu}^2 - \left(m^2 + \frac{6\xi}{R^2}\right) \field_{\ell \mu}^2  \right] \right. \\
\left. -\frac{1}{R^2} (\partial_\pplr \field_{\ell \mu})^2 + \frac{\ell(\ell+1)}{R^2 \sin^2 \pplr} \field_{\ell \mu}^2  \right\} \, .
\end{multline}
The system was thus decomposed into a collection of independent fields $\field_{\ell \mu}(\pplr)$ living on a one-dimensional space, as in \cite{srednicki_entropy_1993}. In contrast with \cite{srednicki_entropy_1993}, however, where the background geometry is that of flat Minkowski spacetime, the one-dimensional space associated with the radial direction is now finite, reflecting the compactness of the spatial sections of the Einstein universe. 

\subsection{Discretization of the model}

\begin{figure}
\centering
\includegraphics{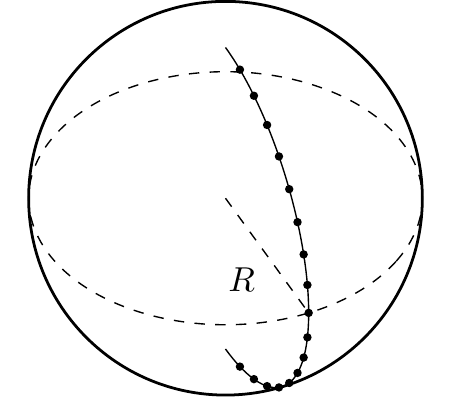}
\caption{Fixing an angle $\plr$, the discretization vertices $\chi_j$ span the range $ 0 \leq \pplr \leq \pi $, covering a physical distance $R \pi = N \varepsilon$. Each vertex is at a distance $ j \varepsilon - N \varepsilon / 2 $ from the equator.}
\label{fig:einsteinspace}
\end{figure}

We now discretize the fields $\field_{\ell \mu}(\pplr)$. The interval $\pplr \in [0,\pi]$ can be partitioned into a union of $N$ subintervals bounded by the equally spaced points
\begin{equation}	\label{eq:def-chi-j}
\pplr_j = \frac{\pi}{N} j \, , \quad j=0,\dots,N \, .
\end{equation}
At any fixed time, each $\pplr_j$ defines a $2$-sphere with area
\begin{equation} \label{eq:area-discrete-entangling}
S_j = 4 \pi R^2 \sin^2 \pplr_j \, .
\end{equation}
The subregions bounded by these surfaces provide a decomposition of $\mathbb{S}^3$ into a union of $N-2$ thick spherical shells and two $3$-balls (at the North and South Poles). The subspace formed by the $n$ first subregions is bounded by a surface of area $S_n$. Denote by $\varepsilon$ the geodesic radial distance between successive boundary surfaces. The maximum distance $ R \pi $ then becomes $ N \varepsilon $ (from the North to the South pole, see Fig.~\ref{fig:einsteinspace}), and we have:
\begin{equation}	\label{eq:R-N}
	R = \frac{ N \varepsilon }{\pi} .
\end{equation}
In addition, the one-dimensional field $\field_{\ell \mu}$ is replaced by a set of $N$ variables that we interpret as living at the center of each subinterval of the decomposition,
\begin{align}
\field_{\ell \mu j} &= \field_{\ell \mu}\left( \pplr_j-\frac{\pi}{2 N} \right)  \, , \nonumber \\
	&= \field_{\ell \mu}( \pplr_{j-1/2}) \, , \quad j=1,\dots,N \, ,
\end{align}
and its partial derivatives can be approximated by finite differences,
\begin{equation}
\partial_\pplr \field_{\ell \mu} \to \frac{\field_{\ell \mu,j+1} - \field_{\ell \mu j}}{\pi/N} \, .
\end{equation}

We now approximate the Lagrangian \eqref{eq:L-ell-mu} by a sum of contributions from each spatial subregion depending on the discretized field. Since the number of independent spatial derivatives is smaller than the number of field variables, terms with spatial derivatives must be handled differently from terms without derivatives in the fields. It is natural to require the discretization to preserve the symmetry under a spatial reflection about the equatorial surface $\pplr = \pi/2$. This can be achieved by discretizing the volume element differently for terms with and without spatial derivatives:
\begin{align} \label{eq:L-terms-discretization}
\sin^2 \pplr \, \field^2_{\ell \mu} \; & \to \; \sin^2 \pplr_{j-1/2} \, \field^2_{\ell \mu j} \, ,  \nonumber \\
\sin^2 \pplr \, (\partial_\pplr \field_{\ell \mu})^2 \; & \to \; \sin^2 \pplr_j \, \left( \frac{\field_{\ell \mu,j+1} - \field_{\ell \mu j} }{\pi/N}  \right)^2 \, .
\end{align}
By doing so, we discretize the integral over $\pplr$ into a sum of $N$ contributions for terms without spatial derivatives, and a sum of $N-1$ contributions for the term including spatial derivatives. In both cases the sum of the discretized volumes gives the total volume of the space:
\begin{align}
\frac{\pi}{2} = \int_0^\pi \dd \pplr \sin^2 \pplr \nonumber 
	&= \frac{\pi}{N} \sum_{j=1}^N \sin^2 \left[ \frac{\pi}{N} \left(j-\frac{1}{2}  \right) \right] \nonumber \\
	&= \frac{\pi}{N} \sum_{j=1}^{N-1} \sin^2 \left( \frac{\pi j}{N}  \right) \, ,
\end{align}
where the integral was replaced by a sum,
\begin{equation}
\int \dd{\pplr} \to \frac{\pi}{N} \sum_j \, .
\end{equation}
The discretized Lagrangian then reads:
\begin{equation}
L = \sum_{\ell \mu} \frac{R^2 \varepsilon}{2} \left[ \sum_{j=1}^N \sin^2 \pplr_{j-1/2} \dot{\field}^2_{\ell \mu j} - \sum_{i,j=1}^N \field_{\ell \mu i} \tilde{V}_{ij} \field_{\ell \mu j} \right] \, ,
\end{equation}
with
\begin{multline}
\tilde{V}_{ij} = \delta_{ij} \left\{ \sin^2 \pplr_{j-1/2} \left[ \frac{\ell(\ell+1)}{R^2 \sin^2 \pplr_{j-1/2}} + m^2 + \frac{6\xi}{R^2} \right] \right. \\
	+ \frac{1}{\varepsilon^2} \left( \sin^2 \pplr_j + \sin^2 \pplr_{j-1} \right)  \\
	\left. - \frac{1}{\varepsilon^2} \left( \delta_{i+1,j} \sin^2 \pplr_i + \delta_{i,j+1} \sin^2 \pplr_j \right) \right\} \, .
\end{multline}

The passage to the Hamiltonian formalism is straightforward. The conjugate momenta are readily obtained,
\begin{equation}
\momt_{\ell \mu j} = \frac{\partial L}{\partial \dot{\field}_{\ell \mu j}} = R^2 \varepsilon \sin^2 \pplr_{j-1/2} \dot{\field}_{\ell \mu j} \, ,
\end{equation}
and we find for the discretized Hamiltonian:
\begin{multline}
H = \sum_{\ell \mu} \left[ \sum_{j=1}^N \frac{1}{2 \varepsilon} \frac{\momt^2_{\ell \mu j}}{R^2 \sin^2 \pplr_{j-1/2}} \right. \\
	\left. +  \sum_{i,j=1}^N \field_{\ell \mu i} \left( \frac{R^2 \varepsilon}{2} \tilde{V}_{ij} \right) \field_{\ell \mu j} \right] \, .
\end{multline}
Applying a canonical transformation that makes the transformed $\field$ and $\momt$ have the same dimensions,
\begin{align}
\field_{\ell \mu j} &\mapsto \frac{\field_{\ell \mu j}}{R \sin \pplr_{j-1/2}} \, , \nonumber \\
\momt_{\ell \mu j} &\mapsto R \sin \pplr_{j-1/2} \momt_{\ell \mu j} \, ,
\end{align}
the Hamiltonian becomes
\begin{align} \label{eq:hamiltonian}
	H &= \sum_{ \one \two } \left(
	\sum_{ j = 1 }^N  \frac{1}{ 2 \varepsilon } \momt_{ \one \two j }^2 +
	\sum_{ i, j = 1 }^N \field_{ \one \two i } V_{ i j } \field_{ \one \two j } \right) \, , \\
	& \equiv \sum_{ \one \two } H_{\one \two} \, ,
\end{align}
with a potential term characterized by the potential matrix
\begin{widetext}
\begin{multline} \label{eq:potential}
	V_{ij} = \frac{ \delta_{ij} }{2 \varepsilon} \left( \varepsilon^2 m^2 +\frac{\pi^2 \ell(\ell+1) }{N^2 \sin^2 \pplr_{i-1/2} } + \frac{\sin^2 \pplr_i}{\sin^2 \pplr_{i-1/2}} + \frac{\sin^2\pplr_{i-1}}{\sin^2 \pplr_{i-1/2}} + \frac{6 \pi^2 \xi }{N^2} \right) \\
	- \frac{1}{ 2 \varepsilon } \left( \delta _{i+1,j} \frac{\sin^2\pplr_i}{\sin \pplr_{i-1/2} \sin \pplr_{i+1/2}} +
	\delta _{i,j+1} \frac{\sin^2 \pplr_j}{\sin \pplr_{j-1/2} \sin \pplr_{j+1/2}} \right) \, .
\end{multline}
\end{widetext}

The discretized Hamiltonian is an explicit sum of decoupled normal modes $\one \two$. Each mode describes a set of coupled harmonic oscillators on a finite one-dimensional lattice. The potential matrix $ V_{ i j } $, which arises from the spatial derivative terms in the original theory in the continuum, is responsible for coupling neighbouring vertices $i$ and $j$. This is the source of entanglement between complementary regions of space, since $V_{ij}$ couples degrees of freedom inside any given region of space to degrees of freedom outside it, which become thus correlated.

%---------------------------------------------------------------------------

\section{Entanglement entropy} \label{sec:method}

\subsection{Universal coefficients: perturbative calculation} \label{sec:perturbative-calculation}

In the continuum, the entanglement entropy of a quantum field is in general divergent, with a leading divergence proportional to the area $\area $ of the surface of the subregion, that is, the entropy satisfies an area law $\entropy \propto \area/\varepsilon^2$, where $\varepsilon$ is an ultraviolet regulator \cite{casini_entanglement_2009}. The proportionality constant depends on the details of the regularization. Corrections to the leading divergence include universal contributions that are independent of the regularization, characterized by the couplings of the theory (see, for instance, \cite{Hertzberg:2010uv,Lewkowycz:2012qr}). The numerical constant appearing in a universal term is called a universal coefficient. 

\begin{figure}
\centering
\includegraphics{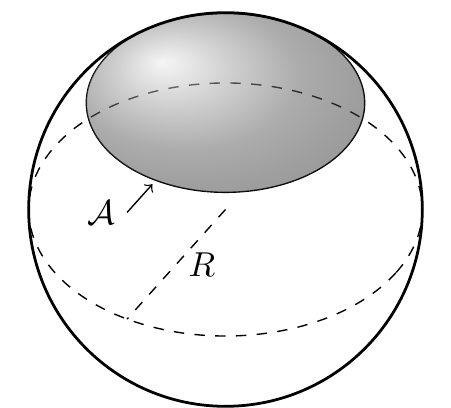}
\caption{\label{fig:subsystem}Circular entangling surface of area $\area$, enclosing the grey region, in the spherical space of radius $R$.}
\label{fig:surface}
\end{figure}

We are interested in computing universal contributions $\entropy_{univ}$ to the entanglement entropy of spatial subregions in the Einstein universe. The spatial sections of constant time for the Einstein universe in the metric \eqref{eq:line} are all isometric to a $3$-sphere $\mathbb{S}^3$ of radius $R$. We consider entangling surfaces $\Sigma$ that are $2$-spheres $\mathbb{S}^2$ of area $\area$ embedded in a constant time slice (see Fig.~\ref{fig:subsystem}). Any such entangling surface splits the spatial $3$-sphere into the union of two complementary subregions, each isometric to a $3$-ball. 

We restrict to the case of a massive field of mass $m$, and consider the regime where correlations in the fluctuations of the field in the complementary subsystems are concentrated in the vicinity of the entangling surface. We thus assume that the correlation length $\lm\sim m^{-1}$  of the scalar field is much smaller than the radius of the entangling surface. The entanglement entropy must then be determined by the form of the metric near the entangling surface, which is a perturbation of a spherical waveguide geometry $ \mathbb{R}^2 \times \mathbb{S}^2 $.

In this regime, the perturbative approach introduced in \cite{Rosenhaus:2014woa,Rosenhaus:2014nha} can be applied to the problem. This approach allows the entropy to be computed as that of the unperturbed metric together with corrections arising from its perturbations. The technique was applied to spherical entangling surfaces in de Sitter space in \cite{Ben-Ami2015}. In this subsection, we review the relevant results of \cite{Ben-Ami2015,Rosenhaus:2014woa,Rosenhaus:2014nha} and apply them to the case of spheres in the Einstein universe. The results so obtained will later be confronted with numerical calculations for the regularization of the unperturbed theory discussed in Section~\ref{sec:model}.

We now proceed to the description of the perturbed spherical waveguide geometry around the entangling surface. Let us introduce a new variable $r$ through:
\begin{equation} \label{eq:arclength}
	\pplr = \frac{\pi}{2} + \frac{r}{R} \, .
\end{equation}
The variable $r \in [-\pi R/2,\pi R/2]$ provides an arc length parametrization along the radial direction with origin at the equator, situated at $\pplr_{eq}=\pi/2$. Following \cite{Ben-Ami2015}, we consider a Wick rotated metric with a temporal coordinate $\tau = it$. The spacetime coordinates are denoted by $x^\mu=(\tau,r, \plr, \azm)$, $\mu=1,\dots,4$. Let $g_{\mu \nu}$ be the Riemannian metric obtained by Wick rotating the line element \eqref{eq:line}. Expanding it around an entangling surface at $r_e$, with $r=r_e + \Delta r$, we find
\begin{align} \label{eq:line-pertb}
	\dd s^2 & = g_{\mu \nu} \dd x^\mu \dd x^\nu \nonumber \\
	& =  \dd \tau^2 + \dd r^2 + R^2 \left[ \cos^2 \left(\frac{r_e}{R}\right) - \sin\left(\frac{2r_e}{R}\right) \frac{\Delta r}{R} \right. \nonumber \\
	& \qquad \left.  - \cos\left(\frac{2r_e}{R}\right)  \left(\frac{\Delta r}{R}\right)^2 + \dots \right] \dd \Omega^2 \, .
\end{align}
The metric near the entangling surface, $\Delta r/R\ll 1$, has the form
\begin{equation}
    g \simeq \bar{g} + h \, ,
\end{equation}
where the background geometry $\bar{g}$ describes a spherical waveguide of radius $R \cos(r_e/R)$, 
\begin{equation}
	\bar{g}_{\mu \nu} \dd x^\mu \dd x^\nu = \dd \tau^2 + \dd r^2 + R^2 \cos^2\left(\frac{r_e}{R}\right)\dd \Omega^2 \, ,
\end{equation}
and the perturbation $h$ is given by
\begin{multline}
	h_{\mu \nu} \dd x^\mu \dd x^\nu = - R^2 \left[ \sin\left(\frac{2r_e}{R}\right) \frac{\Delta r}{R} \right. \\
		\left. + \cos\left(\frac{2r_e}{R}\right)  \left(\frac{\Delta r}{R}\right)^2\right] \dd \Omega^2 \, ,
\end{multline}
up to second order in $\Delta r/R$.

As discussed in detail in \cite{Rosenhaus:2014woa}, the perturbed metric \eqref{eq:line-pertb} can be expressed in terms of the extrinsic and intrinsic curvatures of the entangling surface $\Sigma$. Following their approach, we relabel the coordinates parametrizing the entangling surface as $ y^i = (\plr, \azm) $ and the transverse coordinates as $ x^a = (\tau, \Delta r) $, so that $x^\mu = (x^1, x^2,y^1,y^2)$. We can choose $\tau=0$ for the spatial section of interest, so that both transverse coordinates vanish at $\Sigma$. The unit normals to the entangling surface along the transverse directions have coordinates
\begin{equation}
n_\mu^a = \delta_\mu^a \, , \qquad a =1,2,
\end{equation}
and the metric is such that $g_{ia}=0$ on $\Sigma$, as required in the formalism of \cite{Rosenhaus:2014woa}. We denote the intrinsic metric of the entangling surface by $\gamma = g|_\Sigma$. The line element can then be written in the form:
\begin{multline} \label{eq:line-pertb2}
	\dd s^2 = \left( \delta_{ab} - \frac{1}{3} \mathcal{R}_{acbd} x^c x^d \right) \dd x^a \dd x^b + \left( \gamma_{ij} + \mathcal{R}_{iabj} \, x^a x^b \right. \\
	\left. + 2 K_{aij} x^a + K_{aim} K_b{}^m{}_j \, x^a x^b \right) \dd y^i \dd y^j \, ,
\end{multline}
where the extrinsic curvature is defined as
\begin{equation} 
K^a{}_{ij} = \nabla_i n_j^a \, ,
\end{equation}
and $\mathcal{R}_{\mu \nu \rho \sigma}$ is the curvature tensor of the Einstein universe evaluated at the surface, which has nonzero components only in the purely spatial part:
\begin{equation} \label{eq:curvature-tensor}
\mathcal{R}_{\mu \nu \rho \sigma} = \frac{1}{R^2} (g_{\mu \rho} g_{\nu \sigma} - g_{\mu \sigma} g_{\nu \rho}) \, , \quad \mu,\nu,\rho,\sigma=2,3,4 \, .
\end{equation}
The term $\mathcal{R}_{acbd} x^c x^d$ present in the transverse part of the metric in Eq.~\eqref{eq:line-pertb2} in fact vanishes in the Einstein universe, but we keep it explicitly in the perturbation of the metric in order to obtain results that are valid in a more general class of spacetimes that include, in particular, spheres in the Einstein universe and de Sitter spacetime. For the extrinsic curvature, we find
\begin{equation} \label{eq:extrinsic-curvature}
K^1{}_{ij} = 0 \, , \qquad K^2{}_{ij} = - \frac{R}{2} \sin\left(\frac{2r_e}{R}\right) \bar{\gamma}_{ij} \, ,
\end{equation}
where $\bar{\gamma}_{ij}$ is the metric of the unit two-sphere. Substituting Eqs.~\eqref{eq:curvature-tensor} and \eqref{eq:extrinsic-curvature} in Eq.~\eqref{eq:line-pertb2}, we recover the explicit formula \eqref{eq:line-pertb} for the perturbed metric.

The metric perturbation $h$ can be read from Eq.~\eqref{eq:line-pertb2},
\begin{align} \label{eq:metric-perturbation}
h_{ab} &= - \frac{1}{3} \mathcal{R}_{acbd} x^c x^d \, , \nonumber \\
h_{ai} &= 	0 \, , \nonumber \\
h_{ij} &= \mathcal{R}_{iabj} \, x^a x^b + 2 K_{aij} x^a + K_{aim} K_b{}^m{}_j \, x^a x^b \, ,
\end{align}
and is nonvanishing only tangentially to the surface. It includes contributions from the curvature tensor of the background and from the extrinsic curvature of the entangling surface.

As the starting point for the calculation of the entropy, we consider the unperturbed waveguide geometry. In \cite{Hertzberg:2010uv}, the heat kernel method was used to compute universal mass term contributions to $\entropy$ for a scalar field on a waveguide geometry via the replica trick, for a half-space decomposition with a flat entangling surface. A universal logarithmic term of the form $m^2 \log (m \varepsilon)$ was identified in $(3+1)$-dimensions. On a spherical waveguide, in addition to the universal term identified in \cite{Hertzberg:2010uv}, new universal contributions that are sensitive to the curvature of the entangling surface and background were obtained in \cite{Lewkowycz:2012qr}. We quote their result for the universal terms in $(3+1)$-dimensions \footnote{Eq.~(2.42) of \cite{Lewkowycz:2012qr} with $\alpha=1$, the limit in which the Rényi entropy reduces to the entanglement entropy}:
\begin{equation} \label{eq:univ-bg}
    \entropy^{(0)}_{univ} = \left[ \frac{1}{24\pi} m^2  - \frac{ ( 1 - 6\xi ) }{ 72\pi R^2 \cos^2(r_e/R) } \right] \log(m\varepsilon) \area \, ,
\end{equation}
where we set the radius of the spherical waveguide equal to $R \cos(r_e/R)$. The first term, which is independent of the curvature, agrees with the universal contribution found in \cite{Hertzberg:2010uv}. A different expression for the entropy \eqref{eq:univ-bg} was presented in the Appendix B of \cite{Ben-Ami2015}, however, which follows from a different treatment of the heat kernel, where an extra overall factor of $( 1 - 6\xi )$ appears. In particular, in a conformally coupled theory this universal contribution would then vanish identically. We take Eq.~\eqref{eq:univ-bg} as our formula for the universal terms on the unperturbed geometry, as it will provide a better fit for the numerical results presented later in the paper.

The contribution $ \delta\entropy$ of first order in the metric perturbation $h$ can be calculated using the perturbative approach introduced in \cite{Rosenhaus:2014woa,Rosenhaus:2014nha}, as done on a de Sitter background in the Appendix B of \cite{Ben-Ami2015}. As discussed in \cite{Ben-Ami2015}, such first order contributions have the form:
\begin{equation} \label{eq:delta-S-g-2}
    \delta\entropy = \frac{1}{2} \int \dd^2 x \int \dd^2 y \, \sqrt\gamma \ev{ T^{\mu\nu} K } h_{\mu\nu} \, ,
\end{equation}
where $T^{\mu\nu}$ is the energy-momentum tensor defined as
\begin{equation} 
T_{\mu \nu} = \frac{2}{\sqrt{-g}}\frac{\delta I}{\delta g_{\mu \nu}} \, .
\end{equation}
The integrals in Eq.~\eqref{eq:delta-S-g-2} refer to the transverse and longitudinal directions of the entangling surface in the spherical waveguide background. The average value $\ev{ T^{\mu\nu} K }$ must be computed in the spherical waveguide background. Since we are working in the regime where the correlation length of the field is small compared to the radius of the entangling surface, $\ev{ T^{\mu\nu} K }$ can be approximated by that in a flat geometry, with small curvature corrections. As this quantity is multiplied by the first order perturbation $h$, we can safely neglect the curvature corrections and use the value of $\ev{ T^{\mu\nu} K }$ in flat space, computed in \cite{Cappelli:1990yc}, as done for the case of de Sitter space in \cite{Ben-Ami2015}. From \cite{Rosenhaus:2014ula}, we can write the relevant components of such averages, expressed in our coordinate system, as
\begin{widetext}
\begin{align}
\ev{T_{ij}(r) K} &= - \frac{A_d \, \gamma_{ij}}{(d-1)^2 \Gamma(d)} \int_0^\infty \dd \mu \left[ c^{(0)}(\mu)-(d-1)c^{(2)}(\mu) \right] \mu^2 K_0(\mu r) \, , \nonumber \\
\ev{T_{ab}(r) K} &= - \frac{A_d }{(d-1)^2 \Gamma(d)} \int_0^\infty \dd \mu \left[ c^{(0)}(\mu)+(d-1)(d-2) c^{(2)}(\mu) \right] (\delta_{ab} \mu^2 - \partial_a \partial_b) K_0(\mu r) \, ,
\label{eq:T-spectral}
\end{align}
\end{widetext}
where the spectral functions are given by
\begin{align*}
c^{(0)}(\mu) 	&= \frac{8(d+1)(d-1)}{\Omega_d^2} m^4 \mu^{d-7} \\
			& \quad \times \left( 1-\frac{4m^2}{\mu^2} \right)^{(d-3)/2} \Theta(\mu-2m) \, , \\
c^{(2)}(\mu) 	&= \frac{1}{\Omega_d^2} \mu^{d-3} \left( 1-\frac{4m^2}{\mu^2} \right)^{(d+1)/2} \Theta(\mu-2m) \, ,
\end{align*}
and
\begin{align*}
r &=\sqrt{(x^1)^2 + (x^2)^2} \, , \\
A_d &= \frac{\Omega_d}{(d+1) 2^{d-1} } \, , \qquad 
\Omega_d = \frac{2 \pi^{d/2}}{\Gamma(d/2)} \, . 
\end{align*}
The function $K_0$ is the modified Bessel function of order zero. In our case, the dimensionality of spacetime is $d=4$.

Substituting Eq.~\eqref{eq:T-spectral} and the expression \eqref{eq:metric-perturbation} for the metric perturbation in the formula \eqref{eq:delta-S-g-2} for the entropy, and using the identities
\begin{align*}
\int_{\mathbb{R}^2} \dd^2 x \, x^a x^b K_0(\mu r) &= \frac{4\pi}{\mu^4} \delta^{ab} \, , \\
\int_{\mathbb{R}^2} \dd^2 x \, x^c x^d \partial^a \partial^b K_0(\mu r) &= \frac{2\pi}{\mu^2} (\delta^{ac} \delta^{bd} + \delta^{ad} \delta^{bc}) \, ,
\end{align*}
we obtain an explicit formula for the contributions of the background and extrinsic curvatures for the variation of the entanglement entropy:
\begin{widetext}
\begin{multline} \label{eq:delta-S-g-3}
\delta\entropy = \frac{2 \pi A_d}{(d-1)^2 \Gamma(d)} \int_{\mathbb{S}^2} \dd^2 y \,  \sqrt\gamma \left( \delta^{ac} \gamma^{ij} \mathcal{R}_{iajc}+\frac{1}{2} \delta^{ac} \delta^{bd} \mathcal{R}_{abcd} - \delta^{ab} \gamma^{ij} K_{aim} K_b{}^m{}_j  \right)  \int_0^\infty \frac{\dd \mu}{\mu^2} c^{(0)}(\mu) \\
+ \frac{2 \pi A_d}{(d-1) \Gamma(d)} \int_{\mathbb{S}^2} \dd^2 y \,  \sqrt\gamma \left( \frac{(d-2)}{2} \delta^{ac} \delta^{bd} \mathcal{R}_{abcd} - \delta^{ab} \gamma^{ij} \mathcal{R}_{iajb}+    \delta^{ab} \gamma^{ij} K_{aim} K_b{}^m{}_j  \right)  \int_0^\infty \frac{\dd \mu}{\mu^2} c^{(2)}(\mu) \, ,
\end{multline}
\end{widetext}
A similar expression was obtained in \cite{Ben-Ami2015} for the equatorial surface at the spatial section of minimum radius in de Sitter space. In that case, there are no contributions from the extrinsic curvature, which vanishes at the equator. We further discuss the relation between the results for de Sitter spacetime and the Einstein universe in the Appendix \ref{sec:appendix1}.

The expression \eqref{eq:delta-S-g-3} can be regularized with the introduction of a hard cutoff at $\mu = 1/\varepsilon \equiv \delta$ that eliminates ultraviolet contributions from length scales smaller than $\delta$. The cutoff can then be sent to infinity. The leading divergence comes from the integration of the spectral function $c^{(2)}(\mu)$, which gives:
\begin{equation}
\int_0^\infty \frac{\dd \mu}{\mu^2} c^{(2)}(\mu)  \simeq - \frac{\log(m\varepsilon)}{\Omega_d^2} \, .
\end{equation}
Substituting this leading divergence in \eqref{eq:delta-S-g-3}, integrating over the two-sphere and setting $d=4$, we obtain the universal contribution
\begin{multline}		\label{eq:delta-S-univ-final}
\delta\entropy_{univ} = - \frac{1}{720 \pi}  \left(\delta^{ac} \delta^{bd} \mathcal{R}_{abcd} - \delta^{ab} \gamma^{ij} \mathcal{R}_{iajb} \right. \\
	\left. + \delta^{ab} \gamma^{ij} K_{aim} K_b{}^m{}_j  \right)  \log(m\varepsilon) \area \, .
\end{multline}

We are left with the task of computing the contractions of the curvature tensors,
\begin{align*}
\delta^{ac} \delta^{bd} \mathcal{R}_{abcd} &= 0 \, , \\ 
\delta^{ab} \gamma^{ij} \mathcal{R}_{iajb} &= \frac{2}{R^2} \, , \\
\delta^{ab} \gamma^{ij} K_{aim} K_b{}^m{}_j &= \frac{2}{R^2} \tan^2\left(\frac{r_e}{R} \right) \, ,
\end{align*}
and finally obtain
\begin{equation}	\label{eq:delta-S-univ-formula}
\delta\entropy_{univ} = \frac{1}{360 \pi R^2} \left[1- \tan^2\left(\frac{r_e}{R} \right)  \right]  \log(m\varepsilon) \area \, .
\end{equation}
The first term is a contribution that depends on the background curvature tensor at the entangling surface. The second term is the contribution from the extrinsic curvature. It depends explicitly on the radius $r_e$ of the entangling surface and vanishes at the equator.

The total entropy is the sum of the entropy of the spherical waveguide, given in Eq.~\eqref{eq:univ-bg}, and the contribution from the metric perturbation. Our final result for the universal terms in the entanglement entropy of a scalar field for spheres in the Einstein universe is:
\begin{equation}	\label{eq:S-univ-formula}
\entropy_{univ} = (\alpha_1 m^2 + \alpha_2) \log(m\varepsilon) \area \, ,
\end{equation}
with universal coefficients
\begin{align}	\label{eq:univ-coeffs-formula}
\alpha_1 &= \frac{1}{24\pi} \, , \nonumber \\
\alpha_2 &= - \frac{ ( 1 - 6\xi ) }{ 72\pi R^2 \cos^2(r_e/R) } \nonumber \\
				& \quad + \frac{1}{360 \pi R^2} \left[1 - \tan^2\left(\frac{r_e}{R} \right) \right] \, .
\end{align}

The first universal coefficient $\alpha_1$ is independent of all parameters of the model. It describes a generic subleading logarithmic correction to the area law for the entanglement entropy that is independent of the coupling $\xi$ to the scalar curvature and geometry of the entangling surface and background. It corresponds precisely to the universal term first identified in \cite{Hertzberg:2010uv}. The second universal coefficient $\alpha_2$ can be expressed in terms of the scalar curvature of the background and scalars constructed from the intrinsic and extrinsic curvatures of the entangling surface and that orthogonal to it, as follows.

At the entangling surface, the scalar of curvature $^{(4)}\mathcal{R}$ of the perturbed background geometry is given by
\begin{align}	\label{eq:background-R}
^{(4)}\mathcal{R} &= g^{\mu \rho} g^{\nu \sigma} \mathcal{R}_{\mu \nu \rho \sigma} \nonumber \\
			&= \delta^{ac} \delta^{bd} \mathcal{R}_{abcd} + 2 \delta^{ab} \gamma^{ij} \mathcal{R}_{iajb} + \gamma^{ik} \gamma^{jl} \mathcal{R}_{ijkl} \, .
\end{align}
The tangential components of the curvature tensor are related to the components of the intrinsic and extrinsic curvature of the entangling surface by the Gauss-Codazzi identity \cite{Rosenhaus:2014woa}:
\begin{equation}	\label{eq:gauss-codazzi}
\mathcal{R}_{ijkl} = \prescript{(2)}{}{\mathcal{R}}_{ijkl} + K^a{}_{jk} K_{ail} + K^a{}_{jl} K_{aik}  \, ,
\end{equation}
where $\prescript{(2)}{}{\mathcal{R}}_{ijkl}$ is the intrinsic curvature. An identical relation holds for the transverse surface $\overline{\Sigma}$ parametrized at each point of the entangling surface by the transverse coordinates $x^i$. For a metric of the form \eqref{eq:line-pertb2}, the extrinsic curvature of $\overline{\Sigma}$ vanishes, however, and the Gauss-Codazzi identity reduces to the simpler form:
\begin{equation}	\label{eq:gauss-codazzi-transv}
\mathcal{R}_{abcd} = {}^{(2)}\mathcal{\overbar{R}}_{abcd} \, ,
\end{equation}
where ${}^{(2)}\mathcal{\overbar{R}}_{abcd} $ is the intrinsic curvature of $\overline{\Sigma}$. Eqs.~\eqref{eq:background-R}--\eqref{eq:gauss-codazzi-transv} allow us to express the formula \eqref{eq:delta-S-univ-final} for $\delta S_{univ}$ exclusively in terms of scalars of curvature:
\begin{multline}		\label{eq:S-univ-formula-2}
\delta S_{univ} = -\frac{1}{1440 \pi} \left( - \prescript{(4)}{}{\mathcal{R}} + \prescript{(2)}{}{\mathcal{R}} + 3\, \prescript{(2)}{}{\mathcal{\overbar{R}}} \right. \\
	\left.	+ 3 K^{ail} K_{ail} - K^a K_a  \right) \log(m\varepsilon) \area \, ,
\end{multline}
with $K^a = \gamma^{ij} K^a_{ij}$, and where $ \prescript{(2)}{}{\mathcal{R}}, \prescript{(2)}{}{\mathcal{\overbar{R}}}$ are the Ricci scalars of the entangling and transverse surfaces, respectively,
\begin{align}
\prescript{(2)}{}{\mathcal{R}} &= \gamma^{ik} \gamma^{jl}\, \prescript{(2)}{}{\mathcal{R}}_{ijkl} \, , \\
\prescript{(2)}{}{\mathcal{\overbar{R}}} &= \delta^{ac} \delta^{bd} \, {}^{(2)}\mathcal{\overbar{R}}_{abcd} \, .
\end{align}
As the Gauss-Codazzi identity relates distinct curvature terms, the Eq.~\eqref{eq:S-univ-formula-2} can be equivalently expressed in terms of other sets of independent contractions of the curvature tensors.

The formula for $\delta S_{univ}$ in Eq.~\eqref{eq:S-univ-formula-2} is valid for any metric of the form \eqref{eq:line-pertb2}. In the Einstein universe,
\begin{align}	\label{eq:contractions-Einstein}
\prescript{(4)}{}{\mathcal{R}} &= \frac{6}{R^2} \, , \quad \prescript{(2)}{}{\mathcal{\overbar{R}}} = 0 \, , \nonumber \\
\prescript{(2)}{}{\mathcal{R}} &= \frac{2}{R^2 \cos^2(r_e/R)} \, , \nonumber \\
K^{ail} K_{ail} &= \frac{1}{2}K^aK_a = \frac{2}{R^2} \tan^2\left(\frac{r_e}{R} \right) \, ,
\end{align}
and we recover Eq.~\eqref{eq:delta-S-univ-formula}. The formula can also be directly applied to spheres in de Sitter space, in which case we recover the result of \cite{Ben-Ami2015}, as described in the Appendix \ref{sec:appendix1}. %Eq. (B.7)

In addition, the contribution of the zeroth-order geometry to the universal terms given in Eq.~\eqref{eq:univ-bg} can be expressed in terms of the intrinsic curvature of the spherical entangling surface as
\begin{equation}	\label{eq:univ-bg-2}
    \entropy^{(0)}_{univ} = \left[ \frac{1}{24\pi} m^2  - \frac{ ( 1 - 6\xi ) \prescript{(2)}{}{\mathcal{R}}}{ 144\pi } \right] \log(m\varepsilon) \area \, ,
\end{equation}
The coefficient $1/(24 \pi)$ corresponds to $\alpha_1$, while the second term within the brackets is a contribution to the second universal coefficient $\alpha_2$.

In the special cases of minimal and conformal coupling, the second universal coefficient reduces to
\begin{equation}
\alpha_2 = \begin{dcases}
	\frac{1}{360\pi R^2} \left[ 1 - \frac{5 +\sin^2(r_e/R) }{ \cos^2(r_e/R) } \right] \, , \qquad (\xi=0) \, , \\
	\frac{1}{360 \pi R^2} \left[1 - \tan^2\left(\frac{r_e}{R} \right) \right] \, , \qquad (\xi=1/6) \, .
\end{dcases}
\end{equation}
At the equator, $r_e=0$, the extrinsic curvature vanishes and the formula further simplifies to
\begin{equation}	\label{eq:alpha2-equator}
\alpha_2 = \begin{dcases}
- \frac{1}{90\pi R^2} \, , \qquad (\xi=0) \, , \\
+ \frac{1}{360 \pi R^2}  \, , \qquad (\xi=1/6) \, ,
\end{dcases}
\end{equation}
which also provides a good approximation near the equator, $r_e/R \ll 1$.

In addition to the universal terms, the full entropy of the field includes regularization-dependent terms, both divergent, as the leading contribution to the area law, $\entropy \propto \area/\varepsilon^2$, and finite, as a whole tower of terms involving products of factors of the form $(m \varepsilon)^{2p}$ and $(\varepsilon/R)^{2q}$, with  $p,q \in \mathbb{N}$, multiplied by $\area$, that show up in the integration of the spectral functions during the calculation of $\delta \entropy$.

\subsection{Entanglement entropy in the regularized theory} \label{sec:reg-entropy}

The entanglement entropy of spheres in the Einstein universe can also be computed numerically, exploring the discretization of the theory of a scalar field on this background discussed in Section~\ref{sec:model}. The discretization introduces an ultraviolet regulator at the length scale set by the lattice spacing $\varepsilon$. The entropy of the ground state is then expected to include the universal logarithmic contributions described in Eq.~\eqref{eq:S-univ-formula}. An area law term that scales with $\varepsilon^{-2}$ should also be present \cite{casini_entanglement_2009,bombelli_quantum_1986,srednicki_entropy_1993}, in addition to terms that remain finite in the limit of $\varepsilon \to 0$.

For sufficiently small $\varepsilon$, terms that diverge in the limit $\varepsilon \to 0$ will dominate. In this regime, the universal coefficients $\alpha_1,\alpha_2$ can be determined from calculations of the entropy for different masses. In what follows, we will verify this numerically in order to corroborate our analytical results and, more generally, the perturbative approach developed in the works \cite{Rosenhaus:2014nha,Rosenhaus:2014woa,Ben-Ami2015} and employed in our calculations. We describe the approach adopted for the calculation of the entropy in the discretized theory in this section, and discuss its numerical implementation in the next section.

The system of interest is the canonical quantization of the discretization of the scalar field on the Einstein universe introduced in Section~\ref{sec:model}. The basic observables of the model are the canonical pairs $\{ (\field_{\one \two j}, \momt_{\one \two j} )\}$. The Hamiltonian is given by Eqs.~\eqref{eq:hamiltonian} and \eqref{eq:potential}. It describes a set of coupled oscillators labelled by a multi-index $a=(\one, \two, j)$ over a one-dimensional lattice with nodes $j=1,\dots,N$, with nearest-neighbor interactions described by the off-diagonal components of the potential matrix~\eqref{eq:potential}. The Hilbert space of the system is the tensor product
\[
\mathcal{H} = \bigotimes_{\one \two j} \mathcal{H}_{\one \two j} \, ,
\]
where each $\mathcal{H}_{\one \two j}$ is the Hilbert space of an individual degree of freedom, i.e., the representation space for the canonical pair $(\field_{\ell \mu j}, \momt_{\ell \mu j} )$. The system naturally decomposes into a set of spatially localized subsystems, each associated with a single node $j$:
\[
\mathcal{H} = \bigotimes_j \mathcal{H}_j \, , \qquad \mathcal{H}_j = \bigotimes_{\one \two} \mathcal{H}_{\one \two j} \, .
\]

Let $\mathcal{N}=\{1,\dots,N\}$ be the set of all nodes and $\mathcal{N}_A \subset \mathcal{N}$ a generic subset of nodes. The subsystem $A$ associated with the set of nodes $\mathcal{N}_A$ is described by the Hilbert space
\[
\mathcal{H}_A = \bigotimes_{j \in \mathcal{N}_A} \mathcal{H}_j \, .
\]
It consists of the representation space for the set of canonical pairs with $j \in \mathcal{N}_A$. The complement $B$ of the subsystem is defined analogously with $\mathcal{N}_A$ replaced with its complement $\mathcal{N}_B = \mathcal{N} \setminus \mathcal{N}_A$. We thus obtain a bipartition $\mathcal{H} = \mathcal{H}_A \otimes \mathcal{H}_B$. The subsystem $A$ describes the scalar field restricted to the spatial region formed by the union of the regions associated with nodes in $\mathcal{N}_A$, which are thick spherical shells, for $i\neq 1,N$, or a $3$-ball at the South or North pole, for $j=1$ and $j=N$.

In order to reproduce in the discrete theory the decomposition of $\mathbb{S}^3$ into a union of two glued $3$-balls, we consider a subsystem $A$ formed by the first $n$ nodes. From Eqs.~\eqref{eq:def-chi-j} and \eqref{eq:area-discrete-entangling}, such a subsystem is bounded by an entangling surface of area
\begin{equation}	\label{eq:area-entangling-surface}
 \area = 4\pi R^2 \sin^2 \left( \frac{\pi n}{N}\right) \, .
 \end{equation}
The reduced density matrix is obtained by taking a partial trace over the last $ N - n $ nodes of the full state $\varrho$. The bipartition under consideration has thus the form:
\begin{equation}
	 \underbrace{1, \dots, n}_{\mathclap{\substack{\text{subsystem}\, A}}}, \underbrace{ n+1 , \dots, N }_{\substack{\text{d.o.f.~being} \\ \text{traced out}}} \, .
\end{equation}

We wish to compute the entanglement entropy $\entropy$ of the subsystem $A$ for the ground state of the Hamiltonian \eqref{eq:hamiltonian}. Since the distinct angular momentum modes $(\one, \two)$ are decoupled, they constitute independent subsystems over the one-dimensional lattice with nodes $\mathcal{N}$. The entropy of the subsystem A is then additive over the angular momentum modes,
\[
\entropy = \sum_{ \one \two } \entropy_{ \one \two } \, .
\]
 For a given $(\one, \two)$, the Hamiltonian of the associated mode has the form
\begin{equation} \label{eq:hamiltonian-mode}
	H_{\one \two} = 
	\sum_{ j = 1 }^N  \frac{1}{ 2 \varepsilon } \momt_{ \one \two j }^2 +
	\sum_{ i, j = 1 }^N \field_{ \one \two i } V_{ i j } \field_{ \one \two j }  \, ,
\end{equation}
where the quadratic potential is independent of $\two$. As a result, modes with the same index $\one$ and distinct index $\two$ contribute equally, and we have
\begin{equation} \label{eq:mode-entropy}
	\entropy = \sum_\one ( 2 \one + 1 ) \entropy_\one \, , \qquad \entropy_\one = \entropy_{ \one \two } \, .
\end{equation}
We can then focus on the calculation of $\entropy_{ \one \two }$.

The ground state of a Hamiltonian that is quadratic in the canonical variables is a Gaussian state. General techniques for computing the entropy of Gaussian states are known, and can be directly applied to our problem. A general formula for the entanglement entropy was first derived in \cite{holevo}, reformulated in terms of symplectic invariants in \cite{Adesso:2007} (see also the reviews \cite{Braunstein:2005,Weedbrock:2012,Adesso:2014}), and expressed in terms of the complex structure that characterizes Gaussian states in \cite{bianchi_squeezed_2015} (see the review \cite{hackl2020bosonic}). The case of discretized free field theories has been considered in several works \cite{casini_entanglement_2009,Nishioka:2018khk}, and numerical results have been reported for varied fields and lattices \cite{srednicki_entropy_1993,Lohmayer:2010,Huerta:2012,Casini:2016,huerta2018numerical}. Of particular relevance to our purposes, an efficient algorithm for the numerical calculation of the vacuum entropy of a discrete free field is provided by the real time formalism reviewed in \cite{casini_entanglement_2009,Nishioka:2018khk}. This technique was first applied to the study of entanglement of quantum fields already in the original works where the area law for the entropy was established \cite{bombelli_quantum_1986,srednicki_entropy_1993}. Let us consider the application of the real time approach to the Hamiltonian \eqref{eq:hamiltonian-mode}.

A Gaussian state is completely characterized by the one- and two-point functions of the configuration variables $\field_a$ and their conjugate momenta $\momt_b$, where $a=(\one, \two, i)$. The ground state of a quadratic Hamiltonian has vanishing one-point functions, $\ev{\field_a}=\ev{\momt_a}=0$. Hence, it is sufficient to specify its two-point functions. These are gathered in the covariance matrix $C$, which for vanishing one-point functions has the form:
\[
C = \begin{bmatrix}
\ev{\Phi_a \Phi_b} & \frac{1}{2} ( \ev{\Phi_a \Pi_b} + \ev{\Pi_b \Phi_a}) \\ 
\frac{1}{2} ( \ev{\Pi_a \Phi_b} + \ev{\Phi_b \Pi_a}) & \ev{\Pi_a \Pi_b}
\end{bmatrix} \, .
\]

For a Hamiltonian without mixed terms consisting of products of a field and a momentum operator, the two-point functions of the ground state satisfy the relations
\begin{equation}
\ev{\field_a \momt_b} = - \ev{\momt_b \field_a} = \frac{i}{2} \delta_{ab} \, ,
\end{equation}
so that the off-diagonal blocks of the covariance matrix vanish. The relevant information is then encoded in the symmetric matrices
\begin{equation}
X_{ab} = \ev{\field_a \field_b} \, , \qquad P_{ab} = \ev{\momt_a \momt_b} \, ,
\end{equation}
which in our case take the following form:
\begin{align}	\label{eq:X-P-matrices}
	X_{ i j }^{(\one,\two)} &= \ev{\field_{\one\two i} \field_{\one\two j}} = \frac{1}{2} \left( (2 \varepsilon V)^{-1/2} \right)_{ i j } , \nonumber \\
	P_{ i j }^{(\one,\two)} &= \ev{\momt_{\one\two i} \momt_{\one\two j}} = \frac{1}{2} \left( (2 \varepsilon  V)^{1/2} \right)_{ i j } \, .
\end{align}
See Appendix \ref{sec:appendix2} for details.

Restricting the indices in the correlators to the subset $A$, we obtain the covariance matrix $C_A$ of the subsystem. Denote its diagonal blocks by $X_A$ and $P_A$. All information on observations performed within the subsystem is encoded in the restriction of the covariance matrix to it.  This allows the reduced density matrix to be reconstructed from $C_A$. The entanglement entropy can then be computed from it and expressed directly in terms of the two-point functions. The entropy of the subsystem thus calculated has a simple expression in terms of the positive eigenvalues $\nu_i$ of the matrix $ \sqrt{ X_A P_A } $ \cite{casini_entanglement_2009},
\begin{multline} \label{eq:symplectic-entropy}
	\entropy_{\one \two} = \sum_{i=1}^{n} \left( \nu_i + \frac{1}{2} \right) \log( \nu_i + \frac{1}{2} )  \\
	- \left( \nu_i - \frac{1}{2} \right) \log( \nu_i - \frac{1}{2} ) \, .
\end{multline}
The eigenvalues $\nu_i$ satisfy $\nu_i \geq 1/2$.

This is the formula, together with Eq.~\eqref{eq:mode-entropy}, that we use for the numerical calculations of the entropy. We are thus provided with a convenient shortcut for the calculation of $\entropy$. Instead of taking the partial trace of the full density matrix and using the formula for the von Neumann entropy, which involves traces in a Hilbert space of infinite dimension, it is sufficient to take the first $ n \times n$ entries of the diagonal blocks $X,P$ of the covariance matrix, which defines the matrices $X_A,P_A$, and compute the eigenvalues of $\sqrt{ X_A P_A }$. Such simple manipulations of linear algebra in finite dimensions can be implemented numerically in a straightforward way.

\subsection{Numerical techniques}	\label{sec:numerical-techniques}

The discrete model under consideration has three characteristic length scales: the radius of the universe $R$, the lattice spacing $\varepsilon$ and the inverse mass $m^{-1}$. The entanglement entropy also depends on the choice of the subsystem, which introduces a new length scale $R_e$, the radius of the entangling surface, defined by $\area=4 \pi R_e^2$. We need to fix these parameters for each evaluation of the entanglement entropy. This must be done for an interval of masses respecting the approximations involved in the perturbative calculation of the entropy, and in such a manner that the universal contributions are sufficiently large in comparison with the finite terms so that the universal coefficients can be accurately determined from the numerical results. We consider the case of conformal coupling, $\xi=1/6$. 

The size $N$ of the lattice determines the time required for an individual evaluation of the entropy.  Let us first fix a choice of $N$. In the numerical calculations, an initial value of $N$ will be progressively refined in order to give increasingly accurate estimates of the universal coefficients. A choice of $N$ establishes a relation between two characteristic scales, as the radius of the universe and the lattice spacing are related through $R=N\varepsilon/\pi$, as described by Eq.~\eqref{eq:R-N}. Hence, only one of these length scales is independent, say $\varepsilon$, for a fixed $N$.

For all calculations, we choose the equatorial surface as the entangling surface, in which case $R_e=R$. In the lattice, this can be done by taking an even $N$, and letting the subsystem A be formed by the first $N/2$ nodes. The universal coefficients could be equivalently determined for any other entangling surface, but this choice allows us to avoid introducing an extra length scale in the problem.

We now observe that the entropy is invariant under a certain scaling of the parameters of the model. The existence of such a symmetry can be expected from the fact that the entropy is adimensional. From Eq.~\eqref{eq:X-P-matrices}, for a transformation of the potential matrix of the form $V \to \lambda V$, the components of the covariance matrix transform as $X \to \lambda^{-1/2} X$, $P \to \lambda^{1/2} P$, and the matrix $\sqrt{ X_A P_A }$ whose eigenvalues determine the entropy remains unchanged. Therefore, the entropy is invariant under a transformation $V \to \lambda V$. Moreover, the factors of $2 \varepsilon$ present in $X$ and $P$ as described in Eq.~\eqref{eq:X-P-matrices} cancel in the product $\sqrt{X_A P_A}$ and can be disconsidered for the calculation of the eigenvalues $\nu_i$. In the explicit formula \eqref{eq:potential} for the potential, apart from an overall factor of $1/\varepsilon$, the potential matrix $V$ depends on the length scales only through the combination $\varepsilon m$. Therefore, the entropy is invariant under:
\begin{equation}	\label{eq:rescaling}
\varepsilon \to \lambda \varepsilon \, , \qquad m \to \lambda^{-1} m \, .
\end{equation}
We can then fix the lattice spacing arbitrarily. Results for any $\varepsilon$ can be obtained through the application of the scaling transformation \eqref{eq:rescaling}.

In short, for any given $N$, we have:
\begin{equation}	\label{eq:num-par}
R = \frac{N \varepsilon}{\pi} \, , \quad R_e = R \, , \quad n = \frac{N}{2} \, ,
\end{equation}
where $\varepsilon$ can be fixed arbitrarily, and we consider the case of conformal coupling, $\xi=1/6$. From Eqs.~\eqref{eq:area-entangling-surface} and \eqref{eq:num-par}, the area of the entangling surface is
\[
\area = \frac{4 (N \varepsilon)^2}{\pi} \, .
\]
From Eqs.~\eqref{eq:S-univ-formula}, \eqref{eq:alpha2-equator} and \eqref{eq:num-par}, the universal coefficients determined analytically in the perturbative approach are given for these parameters by:
\begin{align}	\label{eq:universal-coeffs-value}
\alpha_1 &= \frac{1}{24\pi} \quad \simeq \quad 0.0132629 \, ,\nonumber \\
\alpha_2 &= \frac{\pi}{360 (N \varepsilon)^2}  \quad \simeq \quad \frac{8.73 \times 10^{-3}}{(N \varepsilon)^2} \, .
\end{align}
These are the analytical results that we wish to verify numerically. We will keep the product $N \varepsilon$ fixed while refining the lattice to keep the coefficient $\alpha_2$ constant under the variation of the lattice size $N$. We choose an arbitrary unit for lengths and express both $\varepsilon$ and $m^{-1}$ numerically in terms of this unspecified unit.

The calculation of the entropy for a given mass follows three steps. First, for a given angular momentum mode $\one$, the potential matrix is diagonalized and the blocks $X$ and $P$ of the covariance matrix are computed using \eqref{eq:X-P-matrices}. Next, we take the restrictions $X_A$ and $P_A$ and compute the eigenvalues of $ \sqrt{ X_A P_A } $. The entropy of the mode is given by Eq.~\eqref{eq:symplectic-entropy}. Finally, we sum over the modes $\ell$, introducing a cutoff $\one_{max}$, in order to obtain the total entropy. The cutoff $\ell_{max}$ is increased until the numerical fit of the universal coefficients, which we will discuss next, stabilizes.

In addition to the universal terms described in Eq.~\eqref{eq:S-univ-formula}, the entropy includes the non-universal area law term that diverges as $\varepsilon^{-2}$ in the limit of $\varepsilon \to 0$ and is proportional to the area of the entangling surface. Finite terms proportional to $(m \varepsilon)^p \area$ and $(\varepsilon/R)^{q} \area$, $p,q \in 2\mathbb{N}$, are also expected, as discussed in Section \ref{sec:perturbative-calculation}. If $R$ and $\varepsilon$ are kept fixed, the latter finite terms describe a mass-independent contribution proportional to the area $\area$, while the former finite terms become proportional to $m^p \area$. Accordingly, we model the dependence of the entropy on the mass, for fixed $R$ and $\varepsilon$, with the function:
\begin{equation}	\label{eq:ansatz}
\entropy(m) = \left[ \alpha_0 + (\alpha_1 m^2 + \alpha_2) \log (\varepsilon m) + \sum_{\substack{p=2 \\ p \in 2\mathbb{N}}}^{p_{max}} \beta_p \,  m^p \right] \area \, .
\end{equation}
The coefficients $\alpha_r,\beta_s$ can be fitted for a given set of numerical evaluations of the entropy, $\{ \left( \entropy(m_i),m_i \right)\}$. The dependence of the entropy on the unknown coefficients is linear. We estimate them using a multilinear regression based on the least squares method. The maximal power $p_{max}$ of the finite terms can be varied, allowing us to find an optimal choice that minimizes the uncertainties in the numerical estimates of the universal coefficients. If too few finite terms are included in the model, the universal terms absorb contributions of the finite part of the entropy in the fit, which affects the estimation of the universal coefficients, but including too many finite terms can lead to overfitting, making the linear regression more sensitive to numerical noise or systematic errors and thereby increasing the uncertainty in the results.

We are then left with the task of determining a suitable interval of masses to fit the coefficients of the function $\entropy(m)$ given in Eq.~\eqref{eq:ansatz}. In the analytical calculation of the universal coefficients, it is assumed that the correlation length of the field is small in comparison with the radius of the entangling surface, $\lm\sim m^{-1} \ll R_e = R$. In the lattice, this is satisfied if
\begin{equation}	\label{eq:cond-mass-1}
m^{-1} \ll N \varepsilon \, .
\end{equation}
In addition, the lattice must be sufficiently fine so as to provide a reliable approximation of the theory in the continuum. Accordingly, we require the lattice spacing to be small in comparison with the correlation length,
\begin{equation}	\label{eq:cond-mass-2}
\varepsilon \ll m^{-1} \, .
\end{equation}
An optimal value for the mass that takes into account both inequalities on the same footing is determined by the condition
\begin{equation}	\label{eq:optimal-mass}
\frac{1}{mN\varepsilon} = m\varepsilon \quad \Rightarrow \quad m^{-1} = \sqrt{N} \varepsilon \, .
\end{equation}
For sufficiently large $N$, one can expect to find an adequate interval of masses near $m^{-1} = \sqrt{N} \varepsilon$ that allow for a reliable numerical fit of the coefficients in the entropy function \eqref{eq:ansatz}.

To determine such a suitable window of masses, we first computed the entropy for a large set of masses $\mathcal{M}_{scan}=\{m_i\}$ in an interval including the optimal mass $m^{-1} = \sqrt{N} \varepsilon$, and then fitted the curve \eqref{eq:ansatz} for subsets $\{m_i\}_{i=i_{min}}^{i_{max}} \subset \mathcal{M}_{scan}$, varying the position of the mass window, determined by the initial mass $m_{i_{min}}$, and the number $i_{max}-i_{min}+1$ of masses in the windows. This allowed us to determine an interval of parameters where the fit is stable under the variation of the initial mass and the width of the mass window, for which the estimates of the universal coefficients do not change considerably.

The bulk code was written in FORTRAN 90 using the libraries LAPACK \cite{lapack} and OpenBLAS \cite{openblas} as linear algebra solvers, and the package OpenMPI as parallelization framework. For the lattice with $N=1500$ sites, the time required for the calculation of the contribution of each angular momentum mode is of order $\sim11 \, \mathrm{s}/\mathrm{mode}$ on an Intel i3 8100 processor. Our main results were obtained with a cutoff $\ell_{max}=5000$, so that the time required for the calculation of the entropy $S(m_i)$ of a single mass $m_i$ is approximately $15 \, \mathrm{h}$.

%----------------------------------------------------------------------

\section{Numerical results} \label{sec:results}

\begin{table*}
\caption{\label{tab:vary-L} Variation of the estimated universal coefficients under change of the number $\ell_{max}$ of angular momentum modes, determined from a set of 48 masses in the interval $m^{-1}\in (30,50)$ in a lattice with $N=1000$ sites and lattice spacing $\varepsilon=1$. The estimates $\alpha_i^{(1)}$ and $\alpha_i^{(2)}$ are obtained with $\ell_{max}=5000$ and $10000$, respectively. The variation is defined as $\Delta \alpha_i = \alpha_i ^{(2)}-\alpha_i ^{(1)} $.}
\begin{ruledtabular}
\begin{tabular}{ccccc}
  $p_{max}$ & $\alpha_1^{(1)}$  & $\Delta \alpha_1 / \alpha_1^{(1)}$ & $\alpha_2^{(1)}$ &  $\Delta \alpha_2 / \alpha_2^{(1)}$ \\
  	2 & 0.0132295(3) & $-1.90 \times 10^{-6}$    & $-2.0(2) \times 10^{-9}$ & 0.019 \\
	4 & 0.0132595(10) & $-7.83 \times 10^{-5}$  & $8.2(3) \times 10^{-9}$ & -0.046\\
	6 & 0.013245(6) & $-3.27 \times 10^{-4}$ &  $5(2) \times 10^{-9}$ & -0.226  \\
	8 & 0.01336(7) & $8.12 \times 10^{-3}$ &  $2.5(1.2) \times 10^{-8}$  &  0.717 \\
\end{tabular}
\end{ruledtabular}
\end{table*}

We report results obtained for lattices with $N  \geq1000$ sites representing a universe of radius $R=1000/\pi$. Preliminary results indicated that the estimation of the coefficient $\alpha_1=1/24\pi$ can be done accurately in coarser lattices, but not that of the numerically much smaller coefficient $\alpha_2=1/(360 \pi R^2)$, which is strongly affected by errors in the estimation of $\alpha_1$. We observed that the coefficient $\alpha_1$ must be determined with a relative error roughly at the order of $10^{-4}$ in order that $\alpha_2$ can be determined at the percent level, which required the lattices to have at least $N \sim 1000$ sites. We considered the case of conformal coupling and set the entangling surface at the equator, as discussed in Section \ref{sec:numerical-techniques}.

Consider a lattice with $N=1000$ sites and lattice spacing $\varepsilon=1$. From Eq.~\eqref{eq:universal-coeffs-value}, the universal coefficients are then given by
\begin{equation}	\label{eq:R1000-analytical}
\alpha_1 \simeq 0.0132629 \, , \qquad \alpha_2 \simeq 8.73 \times 10^{-9} \, .
\end{equation}
According to Eq.~\eqref{eq:optimal-mass}, the model \eqref{eq:ansatz} should provide a reliable fit of the entropy function $S(m)$ for masses near $m^{-1}=\sqrt{1000} \simeq 31.6$. In order to determine an adequate number $\ell_{max}$ of angular modes in the numerical calculations, we first chose a set of $48$ equidistant masses $\mathcal{M}=\{m_i\}$ in the interval $m^{-1} \in (30,50)$ and studied the variation of the universal coefficients under changes of $\ell_{max}$. We will discuss the choice of the interval of masses in more detail latter.

For a given mass, the contribution of a mode $\ell \mu$ to the total entropy is given by Eq.~\eqref{eq:symplectic-entropy}, where $\nu_i \geq 1/2$. The function
\[
s(\nu_i) = \left( \nu_i + \frac{1}{2} \right) \log( \nu_i + \frac{1}{2} ) 
	- \left( \nu_i - \frac{1}{2} \right) \log( \nu_i - \frac{1}{2} )
\]
that describes the contribution of each eigenvalue $\nu_i$ to $\entropy_{\one \two}$ satisfies
\[
\lim_ {\nu_i \to 1/2} s(\nu_i) = 0 \, ,
\]
so that only for $\nu_i > 1/2$ we have nonvanishing contributions. Numerically, however, the function $s(\nu)$ is not well behaved at $\nu=1/2$; in addition, numerical noise can produce eigenvalues that are numerically less than $1/2$. We first removed such contributions from the calculation of the entropy by introducing a cutoff $\tol$ and restricting the sum \eqref{eq:symplectic-entropy} to include only terms associated with eigenvalues such that $\nu_i-1/2 > \tol$. We decreased the value of the cutoff $\tol$ until no change was observed in the computed entropies $\entropy_{\one \two}$. We verified that this can be attained with $\tol=10^{-35}$. We also checked that the total entropy $\entropy$ obtained by summing over the angular momentum modes as described in Eq.~\eqref{eq:mode-entropy} remained unchanged under further decrease of $\tol$ for $\ell_{max} = 10^4$.

Next we computed the entropy $\entropy(m_i;\ell_{max})$ for several choices of $\ell_{max}$ and estimated the universal coefficients $\alpha_1, \alpha_2$ by fitting the curve \eqref{eq:ansatz} to the numerical data for each choice of $\ell_{max}$. This was done for small values of $p_{max} \in \mathbb{N}$. We observed that for $\ell_{max} \sim 5000$, the relative variation $\Delta \alpha_2/\alpha_2$ in the estimated coefficient with the inclusion of higher $\ell$ modes reached the percent level. The results for $\ell_{max}=5000$ and $10000$ are compared in Table \ref{tab:vary-L}. The relative variation of the coefficient $\alpha_1$ is of the order $\sim 10^{-6}$ for $p_{max}=2$, and rapidly increases with the inclusion of more finite terms in the fit, reaching $\sim 10^{-2}$ for $p_{max}=8$. The relative variation of the coefficient $\alpha_2$ is at the percent level for $p_{max}=2,4$, and increases for larger $p_{max}$, reaching approximately $70\%$ for $p_{max}=8$. The computation time is proportional to $\ell_{max}$. In order to be able to considerably refine the lattice and increase the number of masses in the fit with the available computational resources, we set $\ell_{max}=5000$ for the remaining computations. As a result, our estimation of the universal coefficient $\alpha_2$ will be affected by systematic errors due to the cutoff in $\ell_{max}$ which we estimate to be at the percent level for $p_{max}\leq 4$, and can only provide an order of magnitude estimate for larger values of $p_{max}$. The uncertainties given in Table \ref{tab:vary-L} are the statistical uncertainties in the multilinear regression used in the estimation of the universal coefficients. In these fits, the systematic errors are of the same order of magnitude as the statistical errors in the estimation of the coefficients (except for $\alpha_1, p_{max}=2$, when $\Delta \alpha_1 / \alpha_1^{(1)}$ is even an order of magnitude smaller than the statistical error).

Comparison with the analytical values \eqref{eq:R1000-analytical} shows that a fit with a single finite term, $p_{max}=2$, is inconsistent with the theoretical predictions for $N=1000$. Moreover, for $p_{max}=4$, the estimated values of $\alpha_1$ and $\alpha_2$ are within $4 \sigma$ and $2 \sigma$ from the theoretical predictions, including only statistical errors in the uncertainties. Adding a second finite term thus improves the accuracy of the fit, but the further inclusion of additional finite terms is not advantageous due to the increasing uncertainty in the estimation of the coefficients for fits with more variables, as discussed before. We fix then $p_{max}=4$ for our best estimates of the universal coefficients.

\begin{table}
\caption{\label{tab:refining} Lattice refinement. Universal coefficients estimated from a set of 48 masses in the interval $m^{-1}\in (30,50)$ for lattices with variable number $N$ of sites and fixed size, with $p_{max}=4$.}
\begin{ruledtabular}
\begin{tabular}{ccc}
  N & $\alpha_1$  & $\alpha_2$  \\
  	1000 & 0.0132595(10) & $ 8.2(3) \times 10^{-9}$   \\
	1250 & 0.0132615(14) & $ 8.7(5) \times 10^{-9}$ \\
	1500 & 0.0132625(16) & $ 8.9(5) \times 10^{-9}$
\end{tabular}
\end{ruledtabular}
\end{table}

\begin{figure*}	
\includegraphics[scale=.39]{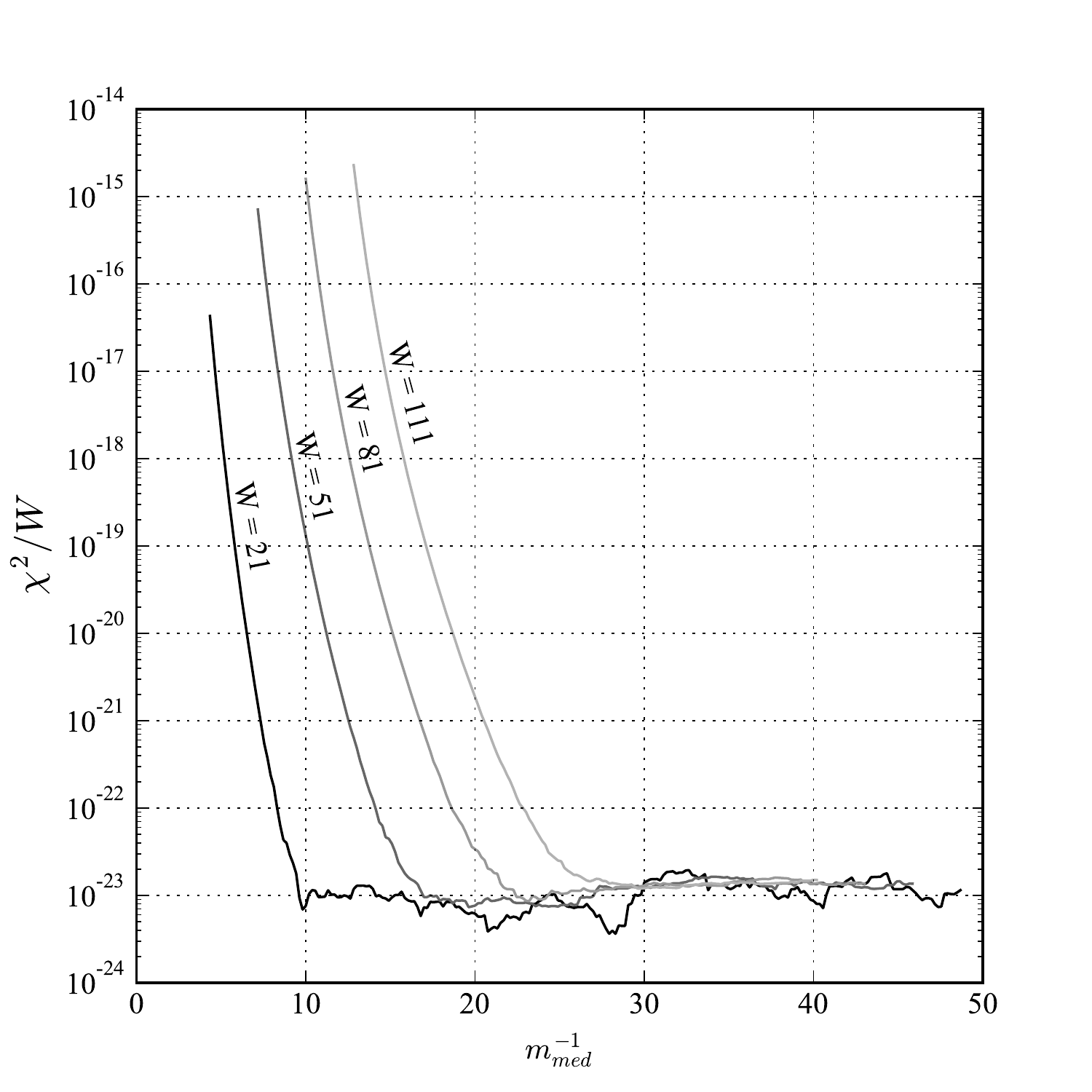}
\includegraphics[scale=.3665]{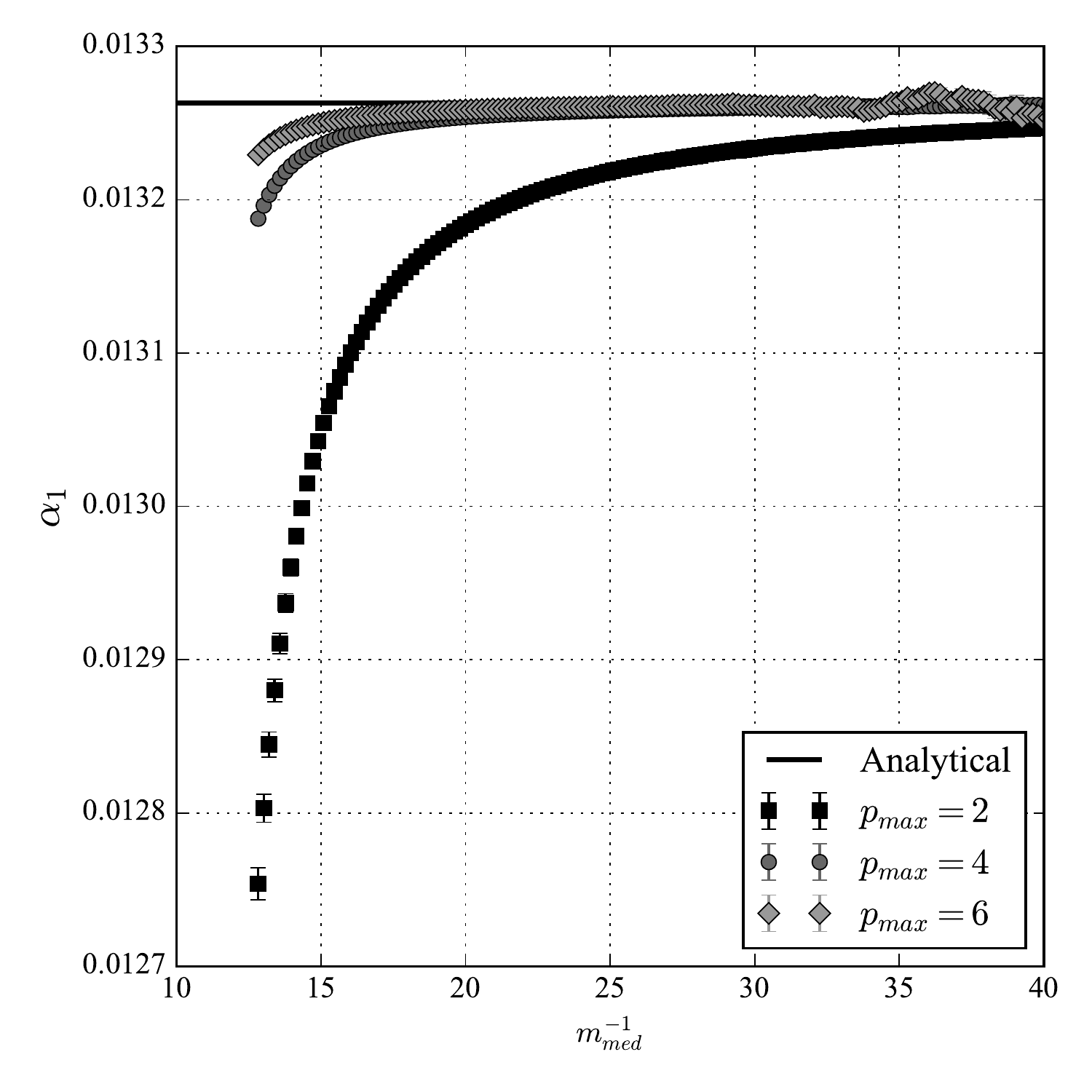}
\includegraphics[scale=.39]{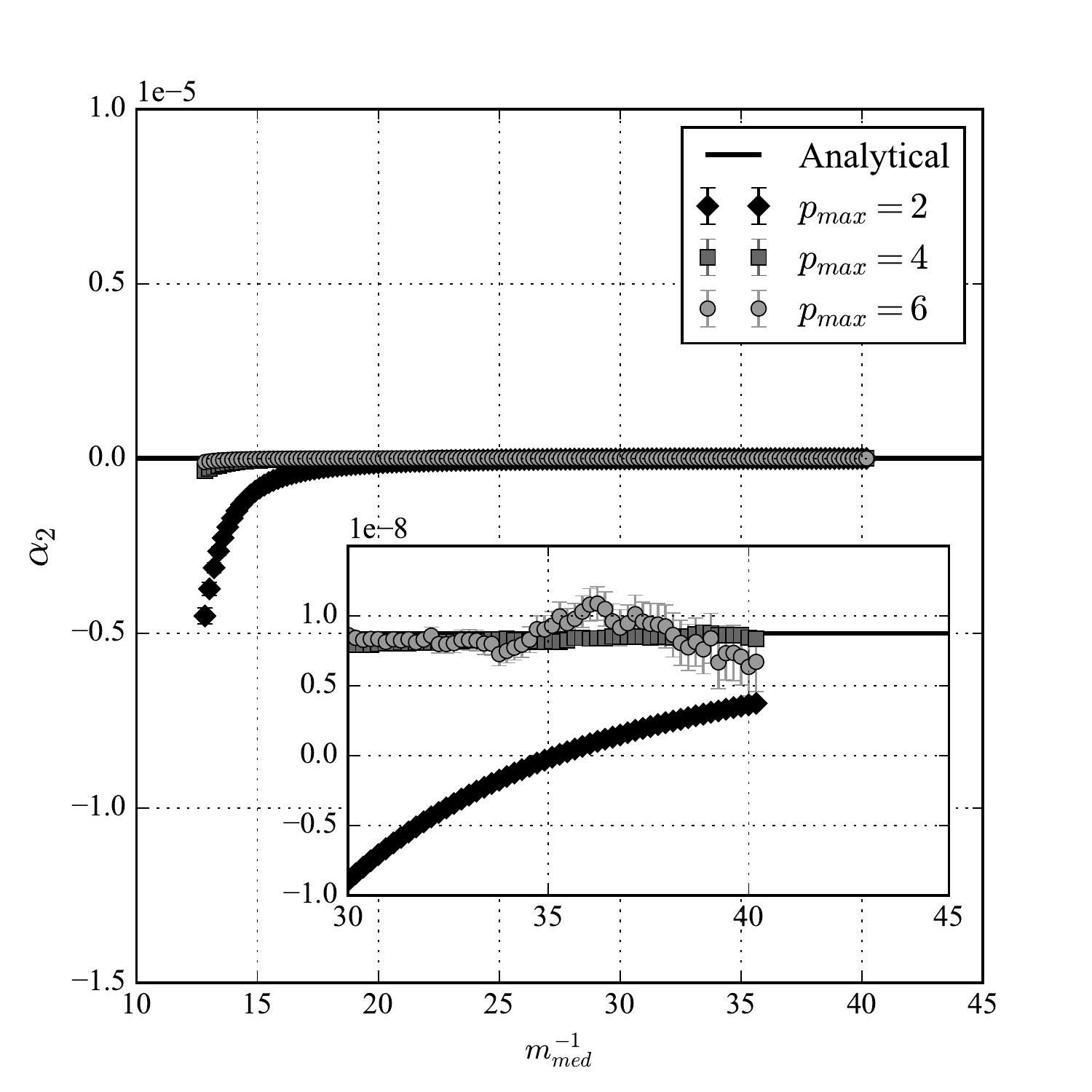}
\caption{\label{fig:chi-squared} \emph{Left panel:} Dependence of the mean squared error per data point, $\chi^2/W$, on the width $W$ and median $m_{med}$ of the mass window. \emph{Middle panel:} Universal coefficient $\alpha_1$ estimated with mass windows of width $W=111$ and median mass $m_{med}$. \emph{Right panel:} Universal coefficient $\alpha_2$ estimated with mass windows of width $W=111$ and median mass $m_{med}$. All results were obtained with the same number of sites, $N=1500$, for a lattice with size $R=1000/\pi$.}
\end{figure*}

Another source of systematic errors is the lattice approximation of the continuum theory. These should decrease by refining the lattice. In finer lattices, the finite terms become less pronounced in comparison with the divergent contributions, and the accuracy of the estimation is expected to improve. With the same set of masses $\mathcal{M}$, we computed the universal coefficients in lattices with a variable number of sites $N=1000,1250,1500$ at fixed $R=N \varepsilon /\pi=1000/\pi$. The results are described in Table \ref{tab:refining}. We found that the coefficient $\alpha_1$ does indeed gradually approach the analytical value given in Eq.~\eqref{eq:R1000-analytical}. The coefficient $\alpha_2$ reaches the analytical value, within its uncertainty, for $N=1250$ and $N=1500$. The estimation of the coefficients is thus stable under refinement of the lattice and approaches the analytical value for both universal coefficients.

From the discussion in Section \ref{sec:numerical-techniques}, for the lattice with $N=1500$ sites and $R=1000$, we expect the fit to be stable in some window of masses near $m^{-1} \simeq 25.8$. In order to verify this, we selected a set $\mathcal{M}_{scan}=\{m_i\}$ of $256$ masses ranging from $m^{-1}=2.45$ to $m^{-1}=50.57$, equally spaced in the axis $m^{-1}$, and fitted the universal coefficients for subsets $\{m_i\}_{i=i_{min}}^{i_{max}} \subset \mathcal{M}_{scan}$, varying the position of the mass window, determined by the initial mass $m_{i_{min}}$, and the number of masses $W=i_{max}-i_{min}+1$. For each mass window, the distribution of inverse masses $m_i^{-1}$ is centered at an inverse mass $m_{med}^{-1}$, where $m_{med}$ is the median of the mass window.

In the first panel of Fig.~\ref{fig:chi-squared}, we plot the mean squared error $\chi^2/W$ for fits with different numbers of masses $W$, for mass windows centered at $m^{-1}_{med}$ in the axis $m^{-1}$. The quantity $\chi^2/W$ is used to evaluate the quality of the fit. We see that it stabilizes for larger $m_{med}$, and reaches a common order of magnitude for all displayed $W$ roughly around $m_{med}^{-1} \sim 25$. At $m_{med}^{-1} \sim 30$, the qualities of all displayed fits are comparable, except for the shorter window $W=20$, which has larger fluctuations in $\chi^2/W$.  At this $m_{med}^{-1}$, the fits involve masses such that $m^{-1} \gtrsim 20$. 

The quantity $\chi^2/W$ stabilizes at larger $m_{med}^{-1}$ for fits involving a larger number of masses $W$. As the inverse masses are equidistant, a fit with larger $W$ involves a larger width $\Delta m^{-1}$ of inverse masses than a fit with smaller $W$. The fits with $W=111$ were performed on intervals of width $\Delta m^{-1} \simeq 20$.  Keeping such a width $\Delta m^{-1} \simeq 20$ fixed, we verified that, for distinct numbers of points $W$ ranging from $21$ to $111$, the quantity $\chi^2/W $  stabilizes roughly at the same $m_{med}^{-1}$ for all $W$. Hence, the tendency observed in Fig.~\ref{fig:chi-squared} of fits with larger $W$ to stabilize at larger $m_{med}^{-1}$ can be attributed to the wider mass window employed in such fits.

We conclude that fits with $W>40$ and masses satisfying $m^{-1} \gtrsim 20$ over an interval of width $\Delta m^{-1} \lesssim 20$ provide estimates of the universal coefficients with a stable quality. In the middle and right panels of Fig.~\ref{fig:chi-squared}, we show the estimated values of the universal coefficients $\alpha_1$ and $\alpha_2$ for mass windows with $W=110$ centered at masses $m_{med}$. We see that the fits are stable for $m_{med}^{-1} \gtrsim 25$ for $p_{max}=4$ and that the values of the estimated universal coefficients agree with the analytical values.

\begin{figure}	
\includegraphics[scale=.5]{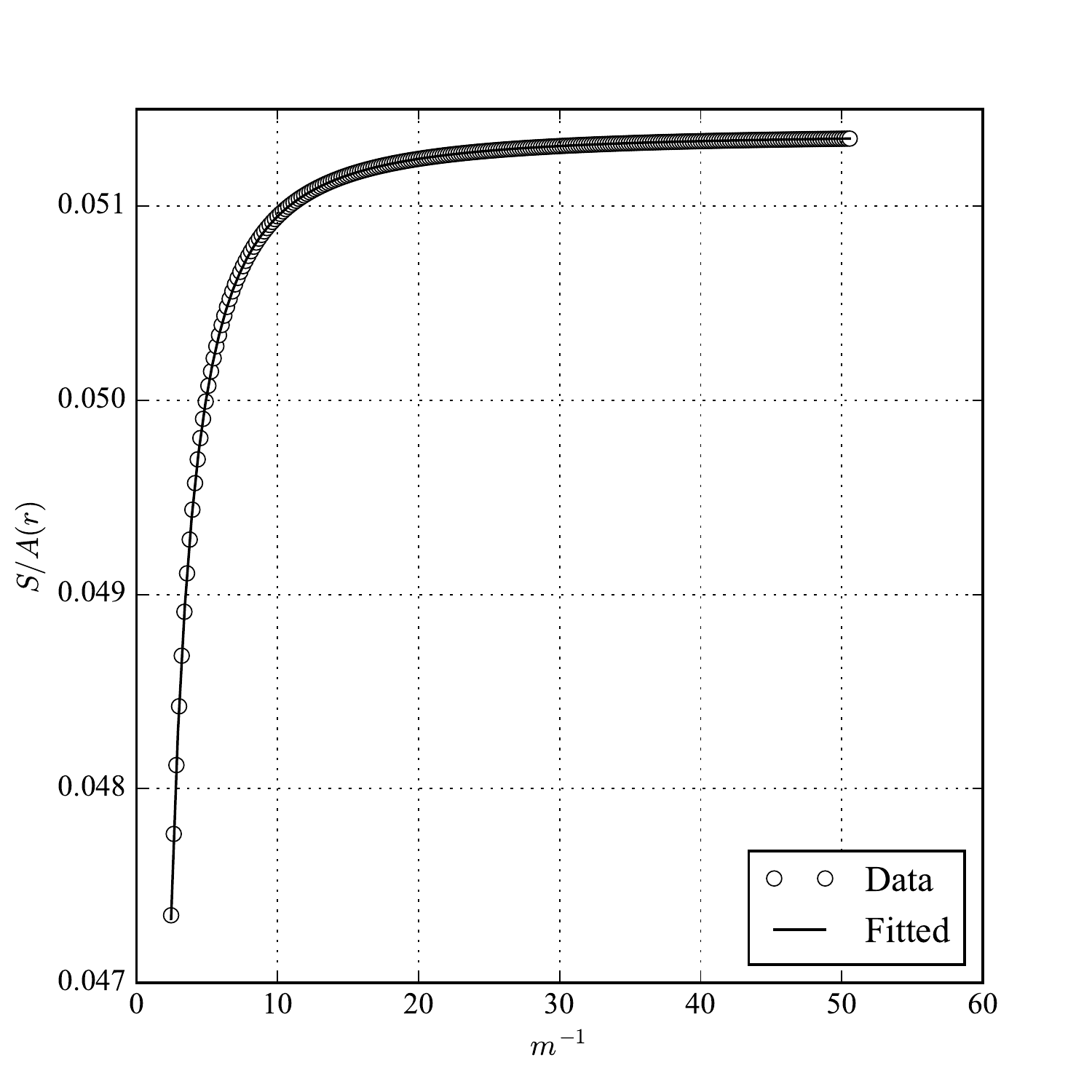}
\caption{\label{fig:entropy-curve} Comparison of the entropy curve $S(m)/A$ with fitted coefficients and the numerical data.}
\end{figure}

In order to obtain our best estimate of the universal coefficients, we performed a fit using $110$ masses such that $m^{-1}>30$ for the lattice with $N=1500$ sites, setting $p_{max}=4$. We obtained:
\begin{align}	\label{eq:num-results}
\alpha_1 &= 0.0132611(11) \, ,  \nonumber \\
\alpha_2 &= 8.43(36) \times 10^{-9} \, .
\end{align}
The coefficient $\alpha_1$ was determined with a relative error of $1.4 \times 10^{-4}$, and the coefficient $\alpha_2$ with a relative error of $3.4 \times 10^{-2}$ with respect to the analytical values \eqref{eq:R1000-analytical}. The uncertainties represented in Eq.~\eqref{eq:num-results} are statistical errors. These are slightly lower than those for the estimates displayed in Table~\ref{tab:refining} for $N=1500$, due to the increased number of masses used in the fit. The error in the estimation of $\alpha_2$ is at the order of the estimated systematic error due to the cutoff $\ell_{max}$ in the sum over angular momentum modes. Further increasing the number of masses in the fit might decrease the statistical error, but in order that the fit provides a more accurate estimation of the coefficients it would be necessary to also decrease the systematic errors, by increasing the number of angular momentum modes in the calculation of the entropy and perhaps further refining the lattice. For the purpose of numerically testing the predictions of the perturbative approach for the calculation of the universal coefficients, we consider that our best estimates already provide strong evidence for the correctness of the entropy formula \eqref{eq:S-univ-formula} with universal coefficients given by Eq.~\eqref{eq:univ-coeffs-formula}, obtained with the application in Section~\ref{sec:method} of the perturbative approach developed in \cite{Rosenhaus:2014woa,Rosenhaus:2014nha}. The first universal coefficient $\alpha_1$ was determined with a relative error of the order $10^{-4}$, and the second, curvature-dependent universal coefficient $\alpha_2$, which to the best of our knowledge has not been obtained numerically before, was determined up to a relative error at the percent level. The fitted entropy function is plotted against the numerical data in Fig.~\ref{fig:entropy-curve}.

%------------------------------------------------------------

\section{Discussion} \label{sec:discussion}

We have determined two logarithmic universal terms of the entanglement entropy for a massive scalar field in four spacetime dimensions, both analytically and numerically. The universal terms are characterized by numerical coefficients $\alpha_1$ and $\alpha_2$,
\begin{equation}	\label{eq:universal-terms}
\entropy_{univ} = \left[ \alpha_1 m^2 \log (\varepsilon m) + \alpha_2 \log (\varepsilon m)\right] \area \, .
\end{equation}
We considered the case of spherical entangling surfaces in the Einstein universe $\mathbb{R} \times \mathbb{S}^3$. The universal coefficients were first computed perturbatively following \cite{Rosenhaus:2014woa,Rosenhaus:2014nha,Ben-Ami2015} and expressed in terms of geometric properties of the background and entangling surface. They were then estimated numerically with the application of the real-time approach \cite{casini_entanglement_2009,Nishioka:2018khk} to a discretization of the theory. We observed a close agreement between the analytical and numerical results.

For the perturbative determination of the universal coefficients, we considered a general class of spherically symmetric spacetimes allowing for the variation of the intrinsic, extrinsic and background geometry at the spherical entangling surface. The geometry around the entangling surface is a perturbed spherical waveguide characterized by the background curvature $R_{\mu \nu \rho \sigma}$ and the extrinsic curvature $K^a{}_{ij}$ evaluated at the entangling surface (Eq.~\eqref{eq:line-pertb2}). Restricting to the case of de Sitter space, we recovered the universal terms described in \cite{Ben-Ami2015}. We focused on the case of the Einstein universe in order to confront the general perturbative formula for the universal coefficients with numerical calculations.

The first universal coefficient $\alpha_1$ is independent of all parameters of the model, $\alpha_1=1/(24 \pi)$, describing a universal term of the entropy characteristic of massive scalar field theories in general geometries and for any coupling $\xi$ to the scalar curvature. It corresponds to the universal term first obtained in \cite{Hertzberg:2010uv}. We determined it numerically with a relative error of the order of $10^{-4}$. Such a high accuracy in the numerical estimation of $\alpha_1$ is required in order that the second universal coefficient $\alpha_2$ can also be estimated from the numerical data. 

 The coefficient $\alpha_2$ includes a contribution from the unperturbed spherical waveguide geometry that depends on the coupling constant $\xi$ to the background scalar curvature and on the intrinsic curvature of $\Sigma$, described by Eq.~\eqref{eq:univ-bg-2}, and contributions from the metric perturbations that include terms proportional to the scalar curvatures $\prescript{(4)}{}{\mathcal{R}},\prescript{(2)}{}{\mathcal{R}},\prescript{(2)}{}{\mathcal{\overbar{R}}}$ of the background, entangling surface $\Sigma$ and the surface $\overbar{\Sigma}$ orthogonal to it, respectively, as well as terms proportional to the contractions $K^{ail} K_{ail}, \, K^a K_a $ of the extrinsic curvature at $\Sigma$, as described in Eq.~\eqref{eq:S-univ-formula-2}. The numerical value of $\alpha_2$ at the equator of the Einstein universe is identical to that for an entangling surface of the same radius at the equator of de Sitter space, computed in \cite{Ben-Ami2015}. For an entangling surface of generic radius, the universal coefficients in the Einstein universe are given by Eqs.~\eqref{eq:S-univ-formula-2} and \eqref{eq:univ-bg-2}, with the required contractions of curvature tensors given explicitly in Eq.~\eqref{eq:contractions-Einstein}. The universal coefficient $\alpha_2$ was computed numerically at the equator of the Einstein universe. Our best estimate agrees with the analytical value up to a relative error of $\sim 3.5\%$. The close agreement between the numerical and analytical results for both universal coefficients provides a stringent numerical test of the perturbative approach to the calculation of the entanglement entropy of massive fields.

For the numerical determination of the universal coefficients, the real-time approach was applied to a lattice model describing a discretization of the scalar field along the radial direction, in a straightforward adaptation of the approach introduced in the original numerical verification of the area law \cite{srednicki_entropy_1993} to the case of the curved background of the Einstein universe and a massive field. The entropy was then computed for a sufficiently large set of masses $\{m_i\}$ and the universal coefficients were obtained by fitting a model for the entropy curve including the universal terms \eqref{eq:universal-terms} to the numerical data $S(m_i)$. Two sources of systematic errors are introduced in this approach: a required cutoff $\ell_{max}$ in the set of angular momentum modes of the field and the lattice regularization itself. These can be reduced by increasing the cutoff $\ell_{max}$ and decreasing the lattice spacing $\varepsilon$, at the cost of increasing the CPU time required for the calculation.  We have progressively improved the numerical calculations in this manner until the estimation of both universal coefficients became reasonably stable. The mass window $\{m_i\}$ was chosen so that the approximations involved in the perturbative calculations were valid. A lattice with $N=1500$ sites, a momentum angular cutoff $\ell_{max} \sim 5000$ and at least $\sim 40$ masses were required for the determination of the coefficient $\alpha_2$ at the percent level. Our best estimate was obtained with a set of $110$ masses and the total CPU time was approximately $\sim \, 1680~\mathrm{h}$.

The numerical approach explored here for the case of a massive scalar field in the Einstein universe can also be applied to the estimation of universal coefficients of the entanglement entropy in other spherically symmetric static geometries and to the case of massless theories. In particular, it should be possible to determine universal terms for a massless scalar field in the Einstein universe with a similar procedure. In the massive case, the entropy curve could also be further studied by extending the range of masses to regimes where the approximations assumed in the perturbative approach do not hold.

%--------------------------------------------------------------------------------

\begin{acknowledgments}
	R. R. S. thanks Prof.~R. Dickman for valuable discussions about the structure of this text, Prof.~M. Smolkin for valuable input concerning the calculations, Mr.~I. Romualdo for discussions leading to better understanding of the topic and Mr.~A. Lara for input in acquiring early results, and acknowledges support from Conselho Nacional de Desenvolvimento Científico e Tecnológico (CNPq) and Coordenação de Aperfeiçoamento de Pessoal de Nível Superior (CAPES) through the Federal University of Minas Gerais and the University of São Paulo.
	
    N.Y. thanks M. Huerta for fruitful discussions during the conference ``VIII Quantum Gravity in the Southern Cone'' and the organization of the event for the support provided. N.Y. acknowledges financial support from the Conselho Nacional de Desenvolvimento Cient\'ifico e Tecnol\'ogico (CNPq) under the grant 306744/2018-0, and from the Programa Institucional de Aux\'ilio \`a Pesquisa de Docentes Rec\'em-Contratados, PRPq/UFMG.
\end{acknowledgments}

%--------------------------------------------------------------------------------

\appendix
\section{Universal coefficients for spheres in de Sitter space}\label{sec:appendix1}

The perturbative calculation of universal terms of the entropy for a spherical entangling surface in a perturbed spherical waveguide was discussed in Section \ref{sec:perturbative-calculation}. We considered the case of the Einstein universe in detail, but a general formula for the universal terms was first derived for any metric of the form \eqref{eq:line-pertb2}. It includes a contribution from the unperturbed spherical waveguide, given by Eq.~\eqref{eq:univ-bg}, and a contribution from the metric perturbation, expressed entirely in terms of scalars of curvature in Eq.~\eqref{eq:S-univ-formula-2}. Here we apply this formula to the case of de Sitter space and show how the results obtained in \cite{Ben-Ami2015} are reproduced from these formulas.

The metric of de Sitter space after a Wick rotation of the temporal variable is that of a four-sphere. It can be written in the form:
\[
ds^2 = R^2 \sin^2 \left( \frac{r}{R}  \right) d\tau^2 + dr^2 + R^2 \cos^2 \left( \frac{r}{R}  \right) d\Omega^2 \, ,
\]
where $R$ is the radius of the four-sphere. Denote by $y^i $ the angular variables parametrizing the two-sphere with metric $d\Omega^2$. Introducing the new variables:
\[
x^1 = r \cos \tau \, , \qquad x^2 = r \sin \tau \, ,
\]
and expanding around the equator ($r=0$), we obtain the perturbed metric,
\begin{multline}	\label{eq:perturbed-metric-dS}
ds^2 = \delta_{ab} dx^a dx^b + \gamma_{ij} dy^i dy^j \\
	+ \frac{1}{3R^2}  \left[ (-x^2)^2 (dx^1)^2 - (x^1)^2 (dx^2)^2 + x^1 x^2 dx^1 dx^2 \right] \\
	- \left[ (x^1)^2 + (x^2)^2  \right]  d\Omega^2 \, ,
\end{multline}
where $\gamma_{ij}$ is the metric of the two-sphere of radius $R$. The first line in Eq.~\eqref{eq:perturbed-metric-dS} corresponds to the unperturbed spherical waveguide, and the remaining terms describe the metric perturbation $h$.

The curvature tensor is given in the full metric by:
\[
\mathcal{R}_{\mu \nu \rho \sigma} = \frac{1}{R^2} (g_{\mu \rho} g_{\nu \sigma} - g_{\mu \sigma} g_{\nu \rho} ) \, .
\]
The extrinsic curvature vanishes at the equator,
\[
K^a_{ij} = 0 \, , \quad \text{ at } x^a=0 \, .
\]
The perturbed metric \eqref{eq:perturbed-metric-dS} is written in terms of the curvature tensor evaluated at the equator as:
\begin{multline} \label{eq:line-pertb3}
	\dd s^2 = \left( \delta_{ab} - \frac{1}{3} \mathcal{R}_{acbd} x^c x^d \right) \dd x^a \dd x^b  \\
	+ \left( \gamma_{ij}  + \mathcal{R}_{iabj} \, x^a x^b \right) \dd y^i \dd y^j \, ,
\end{multline}
which has the form \eqref{eq:line-pertb2} with $K^a_{ij}=0$. This allows us to apply the formula \eqref{eq:S-univ-formula-2} to compute the contribution $\delta \entropy_{univ}$ of the metric perturbations to the universal terms of the entanglement entropy. The contribution from the unperturbed spherical waveguide geometry is the same as before, and given by Eq.~\eqref{eq:univ-bg}.

In de Sitter space, the relevant scalars of curvature are:
\[
\prescript{(4)}{}{\mathcal{R}} = \frac{12}{R^2} \, , \quad \prescript{(2)}{}{\mathcal{R}} = \prescript{(2)}{}{\mathcal{\overbar{R}}} = \frac{2}{R^2} \, ,
\]
and we find
\[
\delta \entropy_{univ} = \frac{1}{360 \pi R^2} \, ,
\]
which corresponds to Eq.~(B.7) of \cite{Ben-Ami2015}. The result is numerically identical to that at the equator of the Einstein universe, but the curvature contributions from the metric perturbations to the result are distinct. At the equator of de Sitter space, the extrinsic curvature vanishes, and the transverse surface has a nonzero intrinsic scalar curvature $\prescript{(2)}{}{\mathcal{\overbar{R}}}$. In the Einstein universe, there are contributions from the extrinsic curvature, while the curvature of the transverse surface vanishes. The contributions from the background are also distinct. In both cases, the entangling surface is a sphere of radius $R$, and the contribution from its intrinsic curvature $\prescript{(2)}{}{\mathcal{R}}$ is the same.

\section{Explicit relation between the covariance matrix and the Hamiltonian}\label{sec:appendix2}

In order to show the relation \eqref{eq:X-P-matrices} between the covariance matrix and the Hamiltonian \eqref{eq:hamiltonian-mode}, we rely on Williamson's theorem, which states that a symmetric, positive-definite matrix can be diagonalized by a linear symplectic transformation $M$.

Let $K$ be the quadratic form defining a quadratic Hamiltonian
\begin{equation}
    H = \frac{1}{2} Z_i K_{ij} Z_j ,
\end{equation}
where $Z$ is a vector in a $2L$-dimensional phase space whose first $L$ coordinates are configuration variables, followed by their conjugate momenta, $Z = (\Phi_a, \Pi_b)$. By Williamson's theorem, there is a symplectic matrix $M$ and a diagonal matrix $D$ such that
\begin{equation}
    K = M D \trp{M} \, .
\end{equation}
This decomposition is referred to as the normal form of $K$.

Following the Appendix A of \cite{bianchi_squeezed_2015}, the symplectic matrix can be evaluated explicitly to be
\begin{equation}	\label{eq:explicit-M}
    M = K^{1/2} U W D^{-1/2} ,
\end{equation}
where
\begin{equation}
    W = \frac{1}{\sqrt{2}}
    \pmqty{ I & \iu I \\ I & -\iu I } ,
\end{equation}
$U$ is a unitary matrix that diagonalizes the matrix $ \iu K^{1/2} J_0 K^{1/2} $,
\begin{equation}\label{eq:will-thm1}
    \iu K^{1/2} J_0 K^{1/2} = U ( \Lambda \oplus -\Lambda ) U^{-1} ,
\end{equation}
where
\begin{equation}
J_0 = \pmqty{ 0 & - I \\ I & 0} \, ,
\end{equation}
$\Lambda$ is diagonal, and $D=\Lambda \oplus \Lambda$, i.e.~the so-called symplectic eigenvalues of $K$ are the positive elements in the spectrum of $\iu K^{1/2} J_0 K^{1/2}$.

In its normal form, the Hamiltonian is that of a collection of free harmonic oscillators; therefore, the covariance matrix of its ground state in the coordinates $Z' = \trp{M} Z$ is $C = I/2$. Under the inverse of this change of coordinates, the symmetric covariance matrix becomes
\begin{equation}
    C = \frac{1}{2} (M \trp{M})^{-1} \, .
\end{equation}

It is sufficient for our purposes to consider a block diagonal $K$ of the form:
\begin{equation}
    K = \pmqty{ A & 0 \\ 0 & c I } \, ,
\end{equation}
with $A$ symmetric and real. Then the matrix $\sqrt{cA}$ has an orthonormal basis of real eigenstates,
\begin{equation}
\sqrt{cA}  u_i = \lambda_i u_i \, .
\end{equation}
Let $u$ be the orthogonal matrix whose $k$-th column is the eigenvector $u_k$, and put $\Lambda = \diag(\lambda_1, \dots, \lambda_L)$, so that
\begin{equation}
\sqrt{cA} = u \Lambda u^T \, .
\end{equation}
It follows that Eq.~\eqref{eq:will-thm1} holds with
\begin{equation}
U = \frac{1}{\sqrt{2}} \pmqty{ u  & u \\ \iu u & - \iu u} \, .
\end{equation}

The symplectic matrix that brings $K$ into its normal form can now be computed using Eq.~\eqref{eq:explicit-M},
\begin{equation}
M = \pmqty{ \sqrt{A} u \sqrt{\Lambda^{-1}} & 0 \\ 0 & - \sqrt{c} u \sqrt{\Lambda^{-1}} } \, ,
\end{equation}
and its inverse is given by
\begin{equation}
M^{-1} = \pmqty{ - \sqrt{\Lambda^{-1}}  u^T \sqrt{c} & 0 \\ 0 & \sqrt{\Lambda^{-1}}  u^T \sqrt{A} } \, ,
\end{equation}
leading to
\begin{equation}
C = \frac{1}{2} \pmqty{ \sqrt{c} A^{-1/2} & 0 \\ 0 & c^{-1/2} A^{1/2} } \, .
\end{equation}
For the case of the Hamiltonian \eqref{eq:hamiltonian-mode}, we set $A = 2V$ and $c = \varepsilon^{-1}$. The resulting covariance matrix is
\begin{equation}
    C = \pmqty{ X & 0 \\ 0 & P }
    = \frac{1}{2} \pmqty{ (2 \varepsilon V)^{-1/2} & 0 \\ 0 & (2 \varepsilon V)^{1/2} } .
\end{equation}

\bibliographystyle{apsrev4-2}

\end{document}